%% file: main_paper.tex
\documentclass{IEEEtran}%, draftclsnofoot,
\usepackage[english]{babel}
\usepackage{blindtext}
\usepackage{tikz,pgfplots}
\usetikzlibrary{calc}
\usepackage{tkz-euclide}
\pgfplotsset{compat=1.5}
\usepgfplotslibrary{polar}
\usepackage{epstopdf} %converting EPS to PDF
\usepackage{amsmath}
\usepackage{wrapfig}
\usepackage{verbatim}
\usepackage{mathrsfs}
\usepackage{amssymb}
\usepackage{accents}
\usepackage{acronym}
\usepackage{graphicx}
\usepackage{textcomp}
\usepackage{xcolor}
\usepackage{mathtools}
\usepackage{color,soul}
\usepackage{graphicx}
\usepackage{booktabs}
\usepackage{multirow}
\usepackage{tikz}
\usepackage[font={small}]{caption}
\usepackage[font={small}]{subcaption}
\usepackage{algorithm}
\usepackage{algorithmic}
\usepackage{fancyhdr}
\usepackage[symbol*]{footmisc}
\usepackage[normalem]{ulem}

\usepackage{cuted}

\usepackage{tabularx}
\usepackage{array} % Required for centering in tabularx

\usepackage{pifont}

\DeclareMathOperator*{\argmin}{arg\,min}

\usepackage{caption}
\usepackage{subcaption}
\usepackage{url}

\newcommand{\eq}[1]{Eq.~\eqref{#1}}

\newcommand{\secref}[1]{Section~\ref{#1}}

\input{ACRONYMS.tex}

\newcommand{\rev}[1]{{\color{black}#1}} %Revised parts
\DefineFNsymbolsTM{otherfnsymbols}{%
	\textdagger    \dagger
	\textsection   \mathsection
	\textdaggerdbl \ddagger
	\textasteriskcentered *
	\textbardbl    \|%
	\textparagraph \mathparagraph
}%

\usepackage{enumitem} %
\usepackage{wrapfig}

\newcommand{\xmark}{\ding{55}} % 

\newcommand{\TO}[1]{\tau_{\mathrm{o},{#1}}} % 
\newcommand{\FO}[1]{f_{\mathrm{o},{#1}}} % 

\newcommand{\textbb}[1]{{\color{black}#1}}

\newcommand{\bbnote}[1]{#1}%{\textbb{(#1)}}

\newcommand{\bb}[1]{\textbb{#1}}

\IEEEaftertitletext{\vspace{-2\baselineskip}}

\begin{document}
	
	\title{Sensing in Bi-Static ISAC Systems with Clock Asynchronism: A Signal Processing Perspective}
	
	\author{Kai~Wu, \IEEEmembership{Member,~IEEE}, 
		Jacopo~Pegoraro \IEEEmembership{Member,~IEEE}, 
		Francesca~Meneghello \IEEEmembership{Member,~IEEE}, \\
  J. Andrew~Zhang, \IEEEmembership{Senior Member,~IEEE}, Jesus~O.~Lacruz, \\
		Joerg~Widmer \IEEEmembership{Fellow,~IEEE}, Francesco~Restuccia \IEEEmembership{Senior Member,~IEEE},\\ Michele~Rossi \IEEEmembership{Senior Member,~IEEE}, Xiaojing~Huang, \IEEEmembership{Senior Member,~IEEE}\\ Daqing~Zhang, \IEEEmembership{Fellow,~IEEE}, Giuseppe~Caire, \IEEEmembership{Fellow,~IEEE}, and Y. Jay~Guo, \IEEEmembership{Fellow,~IEEE}
  \thanks{This work is partially supported by the Australian Research Council under grant No. DP210101411. \\ This work was partially supported by the European Union under the Italian National Recovery and Resilience Plan (NRRP) of NextGenerationEU, partnership on “Telecommunications of the Future” (PE0000001 - program “RESTART”). \\ This work is supported in part by the National Science Foundation under grants CCF-2218845, ECCS-2229472, and ECCS-2329013, by the Office of Naval Research under N00014-23-1-2221 and by the Air Force Office of Scientific Research under grant FA9550-23-1-0261.}
	}

	\markboth{Journal of \LaTeX\ Class Files,~Vol.~14, No.~8, August~2021}%
	{Shell \MakeLowercase{\textit{et al.}}: A Sample Article Using IEEEtran.cls for IEEE Journals}
	
	\newcounter{remark}[section]
	\newenvironment{remark}[1][]{\refstepcounter{remark}\par\medskip
		\textbf{Remark~\theremark. #1} \rmfamily}{\medskip}
	
	\maketitle
	
	\begin{abstract}
		\ac{isac} has been identified as a pillar usage scenario for the impending 6G era. Bi-static sensing, a major type of sensing in \ac{isac}, is promising to expedite \ac{isac} in the near future, as it requires minimal changes to the existing network infrastructure. However, a critical challenge for bi-static sensing is clock asynchronism due to the use of different clocks at far-separated transmitters and receivers. This causes the received signal to be affected by time-varying random phase offsets, severely degrading, or even failing, direct sensing. Hence, to effectively enable \ac{isac}, considerable research has been directed toward addressing the clock asynchronism issue in bi-static sensing. This paper provides an overview of the issue and existing techniques developed in an \ac{isac} background. Based on the review and comparison, we also draw insights into the future research directions and open problems, aiming to nurture the maturation of bi-static sensing in \ac{isac}. 
	\end{abstract}
	
	
	% \textbb{FORMAT REQUIREMENT: up to 20 single column double-spaced pages, 11 point font size, including figures, tables and references. The total number of figures and tables may be up to 10 (sub-figures in (a), (b), (c), etc. counted separately). The total number of references may be up to 30. There should be at least 1.25" margin on left and right sides, and 1" margin from top and bottom. The figures and tables should be placed in the center of the column, and not tightly embedded into the text column. For some special issues, the page limit may be even less if more papers are included. In any event, check with the guest editors before writing the full manuscript.}

	\section{Introduction}
	\label{sec: intro}
	
	% \begin{figure}[!b]
		% 	\includegraphics[width=\linewidth]{./figures/uts/fig_system.png}
		% 	\caption{System diagram \textbb{(will be refined)}}
		% \end{figure}
	
	Communications-centric \acf{isac} systems are expected to deliver a ubiquitous radio sensing network without notably affecting communication performance. These systems have sparked the rise of a novel technological paradigm that seamlessly integrates wireless sensing into next-generation mobile networks. This integration holds the potential to significantly enhance a diverse range of contemporary smart applications~\cite{10024760,8828030, Andrew_PMNsurvey}. A prominent example is the implementation of \ac{isac} in mobile networks, termed the \ac{pmn}. Since its inception in $2017$~\cite{8108564}, \ac{pmn} has captured substantial attention from academia and industry alike. %This intensified interest is palpably reflected by the exponential surge of publications within the open literature dedicated to the topic. %
	An important milestone occurred in June $2023$ when the \ac{itu} greenlighted the 6G vision framework, where the concept of \ac{isac} has been identified as one of the six pivotal usage scenarios for the impending 6G era, also denoted by IMT-2030 \cite{6G_isacUsageScenario}. %This significant announcement has instilled an augmented sense of confidence within the academic and industrial domains, galvanizing collaborative efforts to expedite the development of \ac{isac}.

	\rev{Many sensing approaches are possible in PMN \cite{Andrew_PMNsurvey}, including mono-static sensing at a single base station, bi-static sensing between base stations, or between a mobile device -- \ac{ue} --  and a base station. Bi-static sensing configurations are particularly promising for seamlessly integrating sensing into communications. This approach boasts compelling advantages: It requires almost no change to the current network infrastructure, alleviating the formidable demands of challenging technologies such as full-duplex mobile transceivers. In particular, when the bi-static configuration includes a \ac{ue} as the transmitter, sensing functions benefit from the spatial, spectral, and temporal diversities in user signals, thereby embracing the potency of networked and fused sensing methodologies. Moreover, performing sensing functionalities at the base station is more feasible, as these are equipped with heightened sensing capacities encompassing more advanced power, computational resources, and networking capabilities.} 
	
	% This approach boasts compelling advantages: It requires almost no change to the current network infrastructure, alleviating the formidable demands imposed by challenging technologies such as full-duplex mobile transceivers. In addition, it efficiently leverages the spatial, spectral, and temporal diversities present in user signals, thereby embracing the potency of networked and fused sensing methodologies. Moreover, this methodology is practically feasible, as mobile stations are equipped to accommodate heightened sensing capacities encompassing more advanced power, computational resources, and networking capabilities.
	
	\rev{One critical challenge for bi-static sensing is clock asynchronism due to the bi-static \ac{isac} setup. The bi-static transmitter (Tx) and sensing receiver (Rx), regardless of whether they are base stations or \acp{ue}, are spatially separated, and their clocks from the local oscillators are not locked and asynchronous.} Clock asynchronism causes \ac{to}, \ac{cfo}, and random phase shifts across discontinuous transmissions, which hinder coherent processing of discontinuous channel measurements and introduce ambiguities into sensing results. This is a common problem in almost all communications-centric \ac{isac} systems. Still, it is more prominent in mobile networks due to the typically wide separation of Txs and Rxs, the mobility of nodes, and the absence of line-of-sight connections. Should this problem be efficiently solved, sensing could be seamlessly realized in communication networks via software running in base stations with enhanced computing capabilities. %Communication systems techniques to compensate for phase offsets are \textit{not suitable} for \ac{isac} applications. Indeed, they include the sensing parameters, such as the Doppler shift, into the phase offset and remove them during channel equalization \cite{zhang2022integration, liu2014all}. Initial efforts on \ac{isac} systems have overlooked this problem by assuming radar-like, monostatic full-duplex architectures, or using ad-hoc testbeds with synchronized transmitter-receiver pair.
	
	Considerable research has been directed toward bi-static sensing in \ac{isac} systems to tackle this challenge. These efforts have led to some preliminary solutions, which were partially and conceptually reviewed in an early article \cite{zhang2022integration}. However, these solutions have various limitations, and practically viable solutions are yet to be developed. \rev{In this article, we offer an all-encompassing and meticulous exploration of signal-processing techniques for compensation and removal of \ac{to}, \ac{cfo}, and phase offsets in bi-static sensing. Specifically, this article encompasses \ac{isac} signal processing architectures within contemporary and futuristic mobile networks, which may be applied to any bi-static scenario \textit{without requiring cooperation} or active signaling among nodes (also called \textit{single-node} scenario).} The manuscript is also devoted to thoroughly covering pertinent contributions from the existing open literature, coupled with the sharing of the authors' own reservoir of research experiences, outcomes, and forward-looking visions. \rev{Note that while most wireless sensing literature addresses the clock asynchronism challenge in Wi-Fi networks, the approaches can potentially be applied to mobile networks, given their similar physical-layer techniques. More detailed comparisons between the two networks will be provided in \secref{sec: review bi-static sensing techniques}.} Overall, we hope that our work will contribute to the maturation of the \ac{isac} technology in the near future. 
	
	The rest of this article is organized as follows. In \secref{sec: signal model}, we introduce the different phase offsets caused by clock asynchronism and establish the basic reference \ac{cfr} model used in this paper. A detailed review of three main approaches for eliminating such phase offsets is carried out in \secref{sec: review bi-static sensing techniques}, analyzing and comparing numerous works from the related literature. In \secref{sec:future-work}, we generalize a technical framework and outline six key challenges for future research, focusing on the limitations of the reviewed methods. Finally, we give concluding remarks in \secref{sec:conclusion}.

	\section{Clock Asynchronism and its Modeling in Bi-Static Sensing} \label{sec: signal model}

	\subsection{Offsets Caused by Clock Asynchronism}
The basic clock is typically generated from the \ac{lo} to support various operations in a radio device.
Clock asynchronism happens when the basic clocks of the Tx and Rx are not locked in phase or frequency. 
This introduces unwanted offsets in the received signals -- including but not limited to \ac{cfo}, \ac{to}, and phase offsets -- as detailed next. 
%The sampling offset is typically small when the basic clock is sufficiently stable and accurate, while the first three offsets can have significant impacts on sensing. 
\begin{itemize}
	
	\item \textit{Carrier frequency offset.} \bb{\ac{cfo} is 
		mainly caused by the difference between the \acp{lo} of the transmitter and receiver. 
		Since no two oscillators can generate the same frequency, the carrier frequencies generated by two independent oscillators will also differ, yielding \ac{cfo}. Moreover, due to environmental factors such as temperature changes, supply voltage variations, and aging, oscillator frequency can drift slowly, leading to a slow-varying CFO between a pair of transmitters and receivers.}
	It is typically estimated using, e.g., two segments of repeated signals in communication systems. This estimation can then be used to compensate for the \ac{cfo}. However, the inevitable estimation error leads to residual \ac{cfo}, which may become fast time-varying due to the variation of \ac{cfo} estimation. %For communications, the impact of such residual \ac{cfo} may be ignored or further mitigated via estimating and compensating its accumulated phase shift. However, for sensing, the  residual \ac{cfo} may have a close value to the Doppler shifts to be estimated, causing large estimation error if not being handled properly. 
	
	\item \textit{Timing offset.} \bb{TO results from the lack of synchronized time reference between the transmitter and receiver, as they use their own LOs to generate the necessary timing signals for transmitting and receiving signals.  Due to independent clock sources, there can be an unknown shift or offset in the time perceived by the receiver compared to the actual transmission time. Such TO is almost unchanged during continuous transmission and reception; however, it may become time-varying during discontinuous transmission, particularly when a node transits between transmission and reception. In addition, the change of synchronization point used in the receiver also introduces time-varying TO. The fine timing point may
		slightly vary in position due to, e.g., random noise, even if the channel is unchanged, particularly in
		Orthogonal Frequency Division Multiplexing (OFDM) systems with many subcarriers.
	}
	
	%results from both the asynchronous timing difference between Tx and Rx and the variation of the fine timing point obtained by the signal-synchronization operation at the receiver. \ac{to} is generally fast time-varying in discontinuously received packets/signals. The fine timing point may slightly vary in position due to, e.g., random noise even if the channel is unchanged, particularly in \ac{ofdm} systems with a large number of subcarriers. %For communications, \ac{to} can be absorbed into channel estimation and hence is not an issue at all.

	\item \textit{Phase offset.} \bb{The phase offset can be caused by transceiving electronic devices and the 
		inherent phase noises of \acp{lo}. The factors contributing to these noises, such as thermal noise and flicker noise, are inherently rapid and stochastic, making phase offset fast-varying, possibly on a symbol basis.}

	%phase misalignment between the two device clocks that are too small to be effectively represented by TO. Therefore, it is also often modeled as fast time-varying.

\end{itemize}
\bb{Considering the different contributing factors to the three offsets, we can conclude that: 
	CFO changes on a time scale of minutes to hours due to gradual factors like temperature changes and component aging, but residual CFO after compensation can change on a millisecond time scale; TO changes on a time scale of milliseconds to seconds, as it can be influenced both by slow-varying oscillator drift, transition between transmission and reception, and shift of fine-timing point; and phase offset changes on a time scale of microseconds as it results from rapid, intrinsic noise fluctuations within \ac{lo} and electronic circuits.} 
	
	% For a local oscillator with a stability of 20ppm (parts-per-million), the maximum \ac{to} over 1ms can be 20ns \cite{Andrew_SPoverview_JSTSP}. Comparatively, the propagation delay of a 9-meter path is $9/3e8=30$ ns.
	% Moreover, the \ac{cfo} residual in communication systems can be tens of hertz. For a 5GHz WiFi system, if the \ac{cfo} residual is $100$~Hz, it equivalently introduces a moving target with a radial speed of $ 6$m/s$ (100\times 3\times 10^8/(5\times 10^9)) $. 		
	% To make things worse, the coefficient $\beta_k$ in Eq.~(\ref{eq: Hnk channel matrix}) adds another time-varying phase to the channel matrix. 
	
	% \begin{figure*}
		% 	\centering
		% 	\parbox{0.45\linewidth}{
			% 		\includegraphics[width=\linewidth]{figures/uts/angleCSILTE}
			% 		\caption{Illustrating time-varying phases of LTE's CSIs, as extracted by the NI LTE massive MIMO testbed.}
			% 		\label{fig: CSI phase LTE}
			% 	}\hfill
		% 	\parbox{0.45\linewidth}{
			% 		\includegraphics[width=\linewidth]{figures/uts/angleCSIwifi}
			% 		\caption{Illustrating almost random phases of WiFi's CSIs, as extracted by a \ac{cots} WiFi platform.}
			% 		\label{fig: CSI phase WiFi}
			% 	}		
		% \end{figure*}

\begin{figure}[t!]
		\centering
		%\parbox{0.53\linewidth}{
			\includegraphics[width=0.7\linewidth]{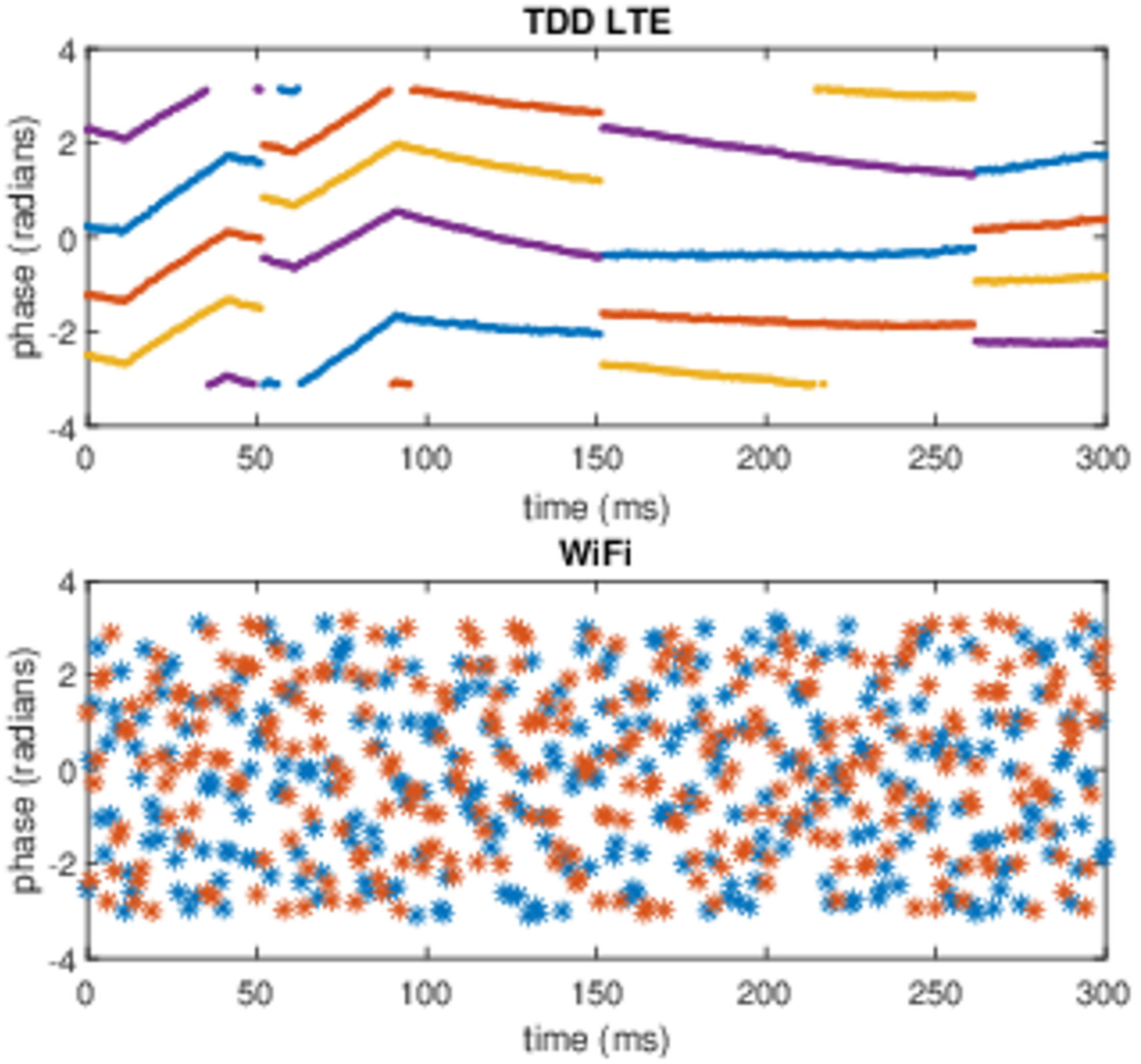}
			\caption{Time-varying phases of \ac{lte}'s (top) and Wi-Fi's (bottom) \ac{csi}s extracted from the National Instrument MIMO testbed and a commercial Wi-Fi platform, respectively. Four different colors are for four equally spaced subcarriers.}
			\label{fig: CSI phase}				
		%}% \hfill
  \end{figure}

	\bb{Fig.~\ref{fig: CSI phase} illustrates the phases of the \ac{csi} estimates over OFDM subcarriers and consecutive symbols for \ac{lte} and Wi-Fi signals in static environments. The phases contain the effects of all offsets mentioned above. For \ac{lte}, which has a stringent timeslot structure, multiple frames (each containing $20$ slots) are transmitted continuously. Each frame starts with a downlink slot, followed by uplink slots. 
		The cell-specific reference signals are used for channel estimation, leading to $40$ \acp{csi} estimation vectors per frame (two symbols with reference signals in each slot). Each vector consists of \ac{csi} estimates over sub-carriers. Four sub-carriers are evenly selected over all resource elements for the illustration in Fig.~\ref{fig: CSI phase}(a).}	 
	%Reference signals are inserted evenly over sub-carriers in each slot for channel estimation. Thus, 
	%	each frame outputs , one for each slot over sub-carriers. The transmission and reception transition happens at the first timeslot of each frame. 
	The figure shows that the \ac{lte} \ac{csi} phase slowly varies over time during continuous transmission, but it suffers from rapid changes corresponding to the uplink-downlink transition. During continuous transmission, phase offset variations are slow as can be seen from the phases of each subcarrier; TO also varies slowly as can be seen from the slow variations of the relative phase differences between the four subcarriers. For Wi-Fi, random packet transmission is performed in the \ac{csi} extraction to emulate practical Wi-Fi scenarios. \bb{For clarity, two sub-carriers (in different colors) are selected for illustration.} We see that the \ac{csi} phase rapidly changes over timeslots and sub-carriers, both in random manners, suggesting fast time-varying phase offset and TO. The results in Fig.~\ref{fig: CSI phase} intuitively manifest the impact of the three offsets illustrated above on bi-static sensing signals. 
	
	%	that these offsets practically exist. 

	\subsection{Signal Modelling}  
	
	%	\begin{enumerate}
		%		\item Illustrate the signal processing architecture in a modern MIMO-\ac{ofdm}/SC transceiver.
		%		
		%		\item Establish the signal model for bi-static sensing in \ac{isac} with practical imperfection factors associated with clock asynchronism, such as clock offset, frequency offset and phase noise. Highlight the different impact of phase offsets on the channel when this is represented in the frequency and delay domains.
		%		
		%		\item Highlight the impact of practical imperfection on bi-static sensing, including intuitive theoretical explanation and experimental results showcased in the literature and from our own research experience.
		%		
		%		\item Explain the challenges in addressing those imperfections, including the restrictions from mobile network systems and signals, signal processing methods and others.
		%		
		%		\item Stress the uniqueness of the clock asynchrony problem in \ac{isac} systems: while synchronization is a well-studied problem in communication, sensing requires a fine grained estimation of the delays and Doppler shifts of each signal propagation path, which are instead considered undesirable offsets to be removed in standard communication systems.
		%		
		%		
		%	\end{enumerate}
	
	We now incorporate these offsets into a \ac{mimo} \ac{ofdm} \ac{isac} signal model. 
	\bb{We note that MIMO-OFDM has been extensively used in modern cellular networks, such as 4G and 5G. It is foreseeable to continue prevailing in 6G.}
	Consider a two-node MIMO-\ac{ofdm} communication system where the transmitting and receiving nodes are each equipped with a uniform linear array. 
	The transmitter (receiver) array has $M$ ($N$) antennas with half-a-wavelength antenna spacing. Their steering vector can be given by 
	%\begin{align}
	$\mathbf{a}(x,\alpha)=[1,e^{j\pi \sin(\alpha)},\dots, e^{j (x-1)\pi \sin(\alpha)}]^{\mathrm{T}},~x=M\text{~or~}N$,
	%\end{align} 
	where $\alpha$ can be the angle of departure (AoD) or arrival (AoA). 
	For \ac{ofdm} transmissions, the frequency band with bandwidth $B$ is divided into $S$ subcarriers. The subcarrier interval is then $\Delta_f=B/S$ leading to an \ac{ofdm} symbol period of $T_O=T_0+T_{cp}$, where $T_0=1/\Delta_f=S/B$ and $T_{cp}$ is the period of the cyclic prefix. 
	%From the perspective of MIMO-\ac{ofdm} communications, the receiver generally starts with timing synchronization to determine the symbol start. Frequency offset is often compensated as well. 
	
	Consider $k=1, \dots, K$ received \ac{ofdm} symbols, equally spaced at an interval of $T_s$. Denote the \ac{to} and (residual) \ac{cfo} as $\TO{k}$ and $\FO{k}$, respectively. 
	Moreover, we use $\beta_k$, a complex value, to reflect random phase offset and all other hardware imperfections (to sensing), such as \ac{agc} and RF chain imbalance, etc.
	\rev{Note that all these nuisances depend on the relative asynchrony between the Tx and the Rx, hence they are specific to each Tx-Rx pair. Moreover, they are time-varying on an \ac{ofdm} symbol level, as indicated by their subscript $k$.} Assume there are $L$ paths, where the $l$-th $(l=1,\dots, L)$ path's complex path gain, propagation delay, Doppler frequency, AoA, and AoD are given by $b_l$, $\tau_l$, $f_{D,l} $, $\phi_{l}$ and $\theta_{l} $, respectively. The noise-free \ac{cfr} matrix at the $i$-th subcarrier and $k$-th \ac{ofdm} symbol can be expressed as 
    %\begin{strip}	
        \begin{align}    		             \mathbf{H}_{i,k}&=\beta_k\sum_{l=1}^L b_l e^{-j2\pi i(\tau_{l}+\TO{k})\Delta_f}
    		e^{j2\pi (f_{D,l}+\FO{k})kT_s}\cdot \notag\\
    		&\quad \quad \quad \quad\mathbf{a}(N,\phi_{l})\mathbf{a}^{\mathrm{T}}(M,\theta_{l})\label{eq: Hnk channel matrix} \\ 
    		&=\underbrace{\beta_ke^{-j2\pi i\Delta_f \TO{k}} 
    			e^{j2\pi \FO{k}kT_s}}_{\xi_{i,k}} 
    		\sum_{l=1}^L b_l e^{-j2\pi i\Delta_f\tau_{l}} \cdot\notag\\    		
      &\quad\quad\quad  \quad  e^{j2\pi f_{D,l} kT_s} 		\mathbf{a}(N,\phi_{l})\mathbf{a}^{\mathrm{T}}(M,\theta_{l}). \label{eq: Hnk channel matrix random phase}
    	\end{align}
    %\end{strip}
 
	Three assumptions have been made here: 1) All transmitting antennas share a common local oscillator, and all receiving antennas do so as well; 2) the intra-symbol phase change caused by Doppler frequency and \ac{cfo} is negligible, and \rev{3) the sensing parameters of the targets can be considered constant within a short coherent processing interval (CPI), such as over a period from several to tens of milliseconds \cite{Andrew_SPoverview_JSTSP}. Thus, their dependence on the symbol index $k$ is suppressed in $\mathbf{H}_{i,k}$ for brevity.} All three assumptions are typical and widely used in most radar and \ac{isac} work. In Eq.~(\ref{eq: Hnk channel matrix random phase}), several nuisance coefficients are combined into a single term, as denoted by $\xi_{i,k}$. 
	\rev{The $L$ paths in Eq. (\ref{eq: Hnk channel matrix random phase}) can be grouped into \textit{static} and \textit{dynamic} paths, depending on whether the underlying scatterer is moving during the observation time period or not. Note that one observation period can consist of multiple OFDM slots or packets. For static paths, we have $f_{D,l}=0$, and $\tau_l, \theta_l, \phi_l$ remain constant over the whole observation period. For a dynamic path $f_{D,l}\neq0$, the sensing parameters can be considered constant within a short processing interval, as done in Eq. (\ref{eq: Hnk channel matrix random phase}). Over the observation period, the sensing parameters may change, as shown later in Section \ref{sec: spatial-domain} and Fig. \ref{fig: CASR xinyu}. These properties are key to some of the methods presented in Section~III.}

	The offsets due to clock asynchronism generally degrade sensing. On the one hand, \ac{to} and \ac{cfo} cause an ambiguity value in delay and Doppler frequency estimation, leading to ambiguity in ranging and speed estimation. On the other hand, the random phase offsets make it impossible to process multiple \ac{csi} measurements coherently directly; note that \ac{csi} power can be exploited for jointly processing multiple measurements in the temporal domain. \bbnote{Moreover, popular advanced sensing techniques from radar sensing, such as the micro-Doppler spectrogram~\cite{chen2006micro}, are also severely affected by \ac{cfo} that destroys Doppler features. The micro-Doppler contains fine-grained information on targets, including multiple moving parts. It is a key component of many applications such as target recognition, human activity recognition, and person identification, among others~\cite{DFRC_radarCentric_mishra2019toward}. We will use the quality of the micro-Doppler as one of the possible evaluation criteria and a means of comparison for \ac{cfo} removal techniques.}

	% \textit{Uniqueness of Clock Asynchronism to \ac{isac}:}
	% One may argue that the clock asynchronism is a classical problem in conventional bi-static radar. This is indeed true, but conventional methods used by radar for addressing the issue can be difficult to apply here. 
	% \begin{itemize}
		% 	\item Bi-static radar, as dedicated for critical sensing tasks, often utilize high-quality clocks, which can be too luxury for everyday or commercial communication devices/infrastructures. As a consequence of average clock stability, the phase offset can fast varying over small time period of milliseconds level, as illustrated in Fig. \ref{fig: CSI phase WiFi}. 
		
		% 	\item GPS-disciplined clocks are typical means used in bi-static radars, which generally target at outdoor sensing scenarios without heavy clutters. 
		% 	This, however, is not difficult to guarantee in \ac{isac}. Utilizing the ubiquitous nature of communication networks, \ac{isac} is aimed to deliver ubiquitous sensing  indoors and outdoors, hence limiting the effectiveness of using GPS-like techniques for addressing the clock asynchronism and suffering from heavy clutters as well. 
		% \end{itemize}

	% \textit{Challenges \bbnote{to be expanded, a single paragraph}:} 1) Complicated propagation environment with rich scattering; 2) \ac{los} or not; 3) Limited number of antennas; 4) Limited time-frequency resources. 	

	% \subsection{Ineffectiveness of traditional offset removal techniques for communication}\label{sec:comm-tech-off}
	
	\subsection{Challenges in Addressing Clock Asynchronism for Sensing} 
	
	In communications systems, the aforementioned offsets due to clock asynchronism are compensated for using well-established algorithms~\cite{liu2014all, schmidl1996low}. Moreover, there is no need to separate TO from the propagation delay for communications. The receiver determines the fine timing, and the remaining timing offsets are included in the channel estimation and removed via channel equalization. %The compensation of \ac{cfo} is also critical, especially in \ac{ofdm} systems to preserve the orthogonality among subcarriers.
	% \ac{cfo} removal methods used in communication systems compensate for the cumulative phase error caused by the \ac{cfo} \textit{plus} Doppler shift, without discerning between the two. This is done similarly in \ac{ofdm} and \ac{sc} systems, following, e.g., \cite{liu2014all, schmidl1996low}. These techniques leverage the repetitions of pilot sequences in the packet preamble to estimate the \ac{cfo} and Doppler shift. The \ac{cfo} plus Doppler-induced phase shift increases \textit{linearly} with time across subsequent samples in the $k$-th symbol/packet preamble as $2 \pi (\FO{k} + f_D)m/B$ where $m$ is the sample index. If the pilot sequence repeats every $M$ symbols, the total phase shift caused by \ac{cfo} and Doppler can be estimated by dividing the phase of the autocorrelation of the received preamble divided by $M$.
	For CFO, it can be estimated and compensated by using, e.g., two repeated sequences \cite{schmidl1996low}, and the impact of residual CFO, together with a small Doppler shift, may be ignored or mitigated via estimating and compensating their accumulated phase shift. %Note that using this method there is no way to separate $f_D$ from $\FO{k}$. 
	
	Unlike data communications, where the offsets can be treated effectively in general, mitigating the impacts of such offsets on sensing is challenging.
	For sensing, the \ac{to} has to be separated to get a clean estimate for propagation delays. The \ac{cfo} and $f_D$ also need to be separated for sensing, and this cannot be achieved via the conventional \ac{cfo} estimation technique. The residual \ac{cfo} may have a close value to the Doppler shifts to be estimated, causing large Doppler estimation errors if not handled properly. Therefore, addressing clock asynchronism is a critical challenge in bi-static sensing.
	
	Several approaches can potentially address this problem \cite{zhang2022integration}, such as using a global GPS disciplined clock (GPSDO) or applying single-node-based signal processing techniques. \bb{A GPSDO with a timing accuracy up to $5.5$~ns can cost $1000$ in US dollars \cite{zhang2022integration}.
		It also requires open-sky view to operate and its size is relatively large. Therefore, it cannot be practically integrated into every mobile device. In comparison, single-node-based processing can be more practical and promising in addressing the asynchronism challenge at a much lower cost, as it does not require additional devices or complex information exchange between multiple nodes. However, GPSDO is not without merits. In scenarios such as a large shopping mall, investing in a high-quality GPSDO may enable the distribution of a precise clock to many terminals via wires, improving clock stability for better bi-static sensing.}		
	This work mainly focuses on signal processing techniques
	for addressing the clock asynchronism issue. Next, we provide an in-depth overview of the techniques.

	% It follows that communication techniques developed to compensate for the offsets are not suitable for sensing, since (i)~a compensation of the aggregated effect of \ac{to} and propagation delays, $\tau_l, \forall l$, does not allow to estimate the distance of the targets, which requires obtaining $\tau_l$ itself, and (ii)~to estimate the target moving speed we aim at removing $\FO{k}$ while \textit{retaining}~$f_D$.

	\section{Single-Node Bi-Static Sensing Techniques } \label{sec: review bi-static sensing techniques}

	% \begin{figure}[!b]
		% 	\centering
		% 	\includegraphics[width=0.6\linewidth]{figures/uts/asyncMethods.pdf}
		% 	\caption{Classification of single node bi-static sensing techniques that tackle the clock asynchronism problem. }
		% 	\label{fig: cacc}
		% \end{figure}
	
\begin{figure}[tb]
  \centering
		%\parbox{0.44\linewidth}{
			\includegraphics[width=0.7\linewidth]{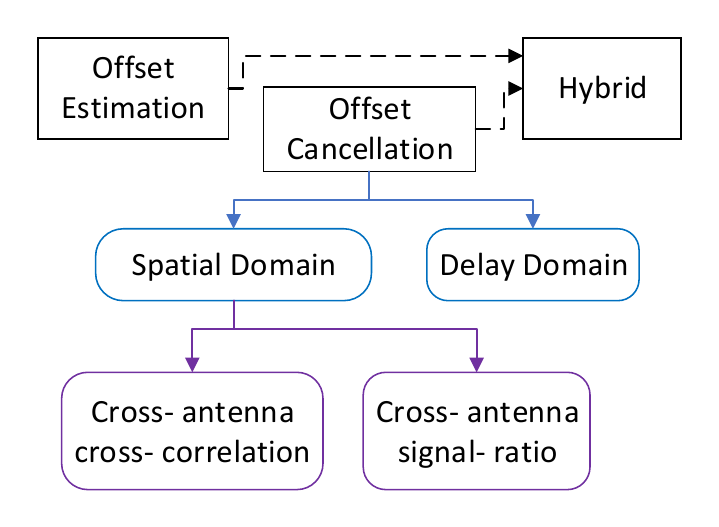}
			\caption{Classification of single node bi-static sensing techniques that tackle the clock asynchronism problem. }
			\label{fig: cacc}			
		%}
	\end{figure}

	From the signal-processing perspective,
	addressing the clock asynchronism issue in bi-static sensing can be achieved by eliminating the contribution of $\xi_{i,k}$ in Eq.~(\ref{eq: Hnk channel matrix random phase}). The elimination can be achieved via either cancellation without estimating the values or estimation followed by compensation; it can also be done in different domains, such as spatial and delay domains. Depending on how the ``elimination'' is achieved, we organize and present the techniques, as found in the literature, through three major types of methods, as summarized in Fig.~\ref{fig: cacc}.

	\begin{enumerate}
		\item Offset Cancellation Methods \cite{CACC,CACC_zhongqin,CACC_ni2020uplink,CASR_farSense,CASR_xinyu,CASR_zhitongTaylor, meneghello2022sharp}: This approach is predominantly applied in the \textit{spatial domain} \cite{CACC,CACC_zhongqin,CACC_ni2020uplink,CASR_farSense,CASR_xinyu,CASR_zhitongTaylor}, utilizing the fact that the clock asynchronism term, $\xi_{i,k}$, is common over antennas and can be mitigated through cross-antenna processing, including cross-correlation and signal ratio. %From Eq.~(\ref{eq: Hnk channel matrix random phase}), it can be seen that either processing can readily remove the term $\xi_{i,k}$. 
		Recently, an offset cancellation in the \textit{delay domain} has been proposed \cite{meneghello2022sharp}, which does not require multiple receiving antennas;  
		
		\item Offset Estimation and Suppression \cite{CASR_multiSense,Uplink_jinbo_TVT, Kai_wifisensing_ICC}: These methods estimate the nuisance terms, namely $\TO{k}$ and $\FO{k}$ separately or their composite impact $\xi_{i,k}$ as an integral, and then mitigate them to facilitate unambiguous sensing;
		
		\item Hybrid of Offset Cancellation and Estimation \cite{pegoraro2023jump}: This type of methods estimate and suppress the \ac{to}, while \ac{cfo} is cancelled out similarly to point 1).  Since the nuisance term, $\xi_{i,k}$, is common to all the propagation paths, delay-domain processing of the \ac{cir} is used to obtain the \ac{to} efficiently through correlation. A reference static path is then used to cancel out the phase offsets without requiring the availability of multiple antennas at the receiver.    	
	\end{enumerate} 
	Each technique has specific features, advantages, and disadvantages, and their applicability depends on the sensing objectives. For example, some techniques may be suitable for obtaining estimates for some specific parameters, such as AoA and delay. Still, they may not be able to enable coherent signal processing over the temporal domains. Next, we review some representative works in the three types in more detail. 
	\rev{Note that the different approaches will be analyzed based on the strategy used to tackle the clock asynchronism and estimate the multi-path propagation parameters, i.e., the paths' complex path gain, propagation delay, Doppler frequency, AoA, and AoD, as described in the CFR model in \eq{eq: Hnk channel matrix random phase}. The target information regarding environmental position and movements can be obtained by applying further processing steps to the estimated propagation parameters. This processing can be independent of the estimation procedure. We refer interested readers to \cite{pegoraro2023jump} for this aspect.}

	\subsection{Offsets Cancellation Methods: Spatial-Domain Processing }\label{sec: spatial-domain}

	\subsubsection{Cross-Antenna Cross-Correlation (CACC) \cite{CACC}} 
	For clarity in reviewing the core idea of \ac{cacc}, we simplify $\mathbf{H}_{i,k}$ in Eq.~(\ref{eq: Hnk channel matrix}) by assuming a single transmitting antenna, i.e., $M=1$. Then the channel coefficient of the $n$-th antenna at the $i$-th sub-carrier and $k$-th \ac{ofdm} symbol can be written into
	\begin{align}\label{eq: h_ikn}
		H_{i,k,n}=\xi_{i,k}	
		\sum_{l=1}^L b_l e^{-j2\pi i\Delta_f\tau_{l}}
		e^{j2\pi f_{D,l}kT_s}e^{j n\pi \sin(\phi_l)}.
	\end{align}
	By name, \ac{cacc} takes the conjugate cross-correlation of the channel coefficients between two antennas. One antenna is considered as a \textit{reference} and given index $n=0$ by convention. 
	Let us consider the cross-correlation between $H_{i,k,0}$, i.e., $n=0$, with $H_{i,k,n}$, where $n\ne 0$. The \ac{cacc} based on $H_{i,k,n}$ can be calculated as
 
%\begin{strip}	
 \begin{align}\label{eq: r_ikn}
		r_{i,k,n} &= H_{i,k,0}H_{i,k,n}^* \notag\\
  &\approx  \sum_{l_1=1}^L\sum_{l_2=1}^L b_{l_1}b_{l_2}^* e^{-j2\pi i\Delta_f(\tau_{l_1}-\tau_{l_2})}\cdot \notag\\ 
  &\quad\quad e^{j2\pi (f_{D,l_1}-f_{D,l_2} )kT_s}e^{-j\pi n \sin(\phi_{l_2})}, 
	\end{align}
 %\end{strip}
\noindent where the approximation is due to $|\xi_{i,k}|^2\approx 1$. The phase offset is fully mitigated from $r_{i,k,n}$. However, its cross-products also introduce more unknown parameters by creating virtual paths. 
	Numerous designs have been proposed in the literature to efficiently estimate the true target parameters from $r_{i,k,n} $~\cite{CACC,CACC_zhongqin,CACC_ni2020uplink}. 
	
	In~\cite{CACC,CACC_zhongqin}, a strong \ac{los} is assumed to be present, a single dynamic path is considered, and the $L$ paths in Eq.~(\ref{eq: h_ikn}) are divided into two groups: static paths and a single dynamic path. Then $r_{i,k,n}$ in Eq.~(\ref{eq: r_ikn}) has four main components:
	\begin{enumerate}
		\item Static paths on antenna $n=0$ times the conjugate of static paths on antenna $n$: in the presence of a strong \ac{los}, this product is approximately constant over time and can be suppressed using a high-pass filter. 
		
		\item Dynamic path on the antenna $n=0$ times the conjugate of the dynamic path on the antenna $n$: This product is much weaker than the other three involving the strong \ac{los}-dominant static path, and hence can be ignored. 
		
		\item Static paths on the antenna $n=0$ times the conjugate of the dynamic path on the antenna $n$: This product includes the dynamic path's sensing parameters, which have opposite signs to actual ones due to the conjugate operation. Its power is also dominated by the static path on antenna $0$.
		
		\item The dynamic path on antenna $0$ times the conjugate of static paths on antenna $n$: This product reflects true sensing parameters with power dominated by the static paths on antenna $n$. 
		
	\end{enumerate}

	Based on the above four components' features, an add-minus sensing method (AMS) is proposed in \cite{CACC}. As can be seen from the power features of the last two components above, adding a positive value to the antenna $n$ can enhance the last component, and subtracting a positive value from antenna $0$ can weaken the third component with the virtual target. This then facilitates the estimation of the actual target's sensing parameters based on \ac{cacc}. Unlike AMS, the work in \cite{CACC_zhongqin} proposes utilizing the other components listed above to mitigate unwanted terms. 
	In particular, the authors of \cite{CACC_zhongqin} reveal that, in the presence of a strong \ac{los} path, the second-order cyclic differences of all cross-correlations over adjacent antennas can be used for estimating the terms related to the virtual targets in the first-order difference. The estimation can then be used to mitigate the impact of virtual targets in the first-order differences, facilitating the unambiguous Doppler frequency estimation using the root-MUSIC algorithm.

	Different from the above work focusing on a single dynamic path, the work in \cite{CACC_ni2020uplink} constructs new signals based on \ac{cacc} results in such a way that the constructed signals are spanned by the same number of bases as that of the dynamic paths. A mirrored MUSIC algorithm is then proposed to estimate the Doppler frequencies of multiple targets. 
	
	If multiple snapshots are exploited to generate AoA estimates only, we can compute the conventional spatial correlation matrix $\frac{1}{K}\sum_{k=k_0}^{k_0+K-1}\mathbf{H}_{i,k}\mathbf{H}_{i,k}^H$, where all offsets are canceled as CACC is implicitly applied. Thus, conventional spectrum analysis techniques such as MUSIC can be applied to obtain the estimates. However, it is important to note that the cross-correlation terms from $K$ snapshots are averaged as the output, preventing the application of a bandpass filter to eliminate static components. Fortunately, as shown in \cite{Uplink_jinbo_TVT}, all static components can be estimated as one element in the MUSIC spectrum domain, as they appear as one constant vector in all snapshots.
	
	\subsubsection{Cross-Antenna Signal Ratio (CASR) \cite{CASR_farSense}}\label{sec:CASR} 
	Similar to \ac{cacc}, \ac{casr} utilizes the fact that the phase offset caused by asynchronism is common to all receiving antennas, the ratio between two antennas can cancel out the asynchronism factors, i.e., $\xi_{i,k}$ in Eq.~(\ref{eq: Hnk channel matrix random phase}), facilitating unambiguous sensing. 
	Similar to how we obtain $r_{i,k,n}$ in Eq.~(\ref{eq: r_ikn}), the ratio between $H_{i,k,n}$ and $H_{i,k,0}$ provides the following \ac{casr} formulation
	\begin{align}
		\rho_{i,k,n} = \frac{H_{i,k,n}}{H_{i,k,0}} = \frac{\sum_{l=1}^L b_l e^{-j2\pi i\Delta_f\tau_{l}}
			e^{j2\pi f_{D,l}kT_s}e^{j n\pi \sin(\phi_l)}}{ \sum_{l=1}^L b_l e^{-j2\pi i\Delta_f\tau_{l}}
			e^{j2\pi f_{D,l}kT_s} }, 
	\end{align}
	where $\xi_{i,k}$ is canceled out, but the sensing parameters appear in the denominator. The latter change makes \ac{casr} a non-linear model of sensing parameters, invalidating numerous conventional sensing methods based on a linear sensing model.  
	
	Assuming a single dynamic path and combining all static paths into a single term, the \ac{casr} formulation above can be simplified as 
	\begin{align}\label{eq: rho_ikn}
		\rho_{i,k,n} = \frac{ \tilde{b}e^{j n\pi \sin(\phi_l)}e^{j2\pi f_{D,l}kT_s} + \mathcal{A}  }{ \tilde{b} e^{j2\pi f_{D,l}kT_s} + \mathcal{B} }, 
	\end{align}
	where $\mathcal{A}$ and $\mathcal{B}$ are independent of $ f_{D,l} $. It is pointed in \cite{CASR_farSense} that, if $\mathcal{A} \tilde{b} e^{j2\pi f_{D,l}kT_s}-\mathcal{B}  \tilde{b}e^{j n\pi \sin(\phi_l)} \ne 0 $, $\rho_{i,k,n}$ is the Mobius transformation of $e^{j2\pi f_{D,l}kT_s} $. Through analyzing the different impact of translation, complex inversion, and multiplication
	operations involved in a Mobius transformation, it is revealed that the \ac{casr} and $e^{j2\pi f_{D,l}kT_s} $ span the same angular range in the complex domain. In the case of a single dynamic path, as considered in \cite{CASR_farSense}, the absolute \ac{casr} 
	presents a sinusoidal-like oscillation and the peak-to-peak magnitude of the oscillation can be enhanced by rotating the \ac{casr} in the complex domain. After searching for the optimal 
	rotation, the autocorrelation of the absolute \ac{casr} over symbols is used to estimate the dynamic path's Doppler frequency.

	In \cite{CASR_xinyu}, three estimation methods are proposed to estimate the Doppler frequency of a single target based on the \ac{csi} ratio. 
	The first method estimates the angular change of $\rho_{i,k,n}$ over a time interval, say $\Delta_t$, and then performs the linear fitting to estimate the changing rate of the phase as the Doppler frequency. 
	Note that, to estimate the angular change over time, the center of the arc/circle generated by $\rho_{i,k,n}$ must be estimated first. The second method is built on the same periodicity between $\rho_{i,k,n}$ and $e^{j2\pi f_{D,l}kT_s} $. It proposes to estimate the periodicity of $\rho_{i,k,n}$ over $k$ by searching for the time interval leading to the closest angles of $\rho_{i,k,n}$.  
	The third method utilizes the periodicity of $\rho_{i,k,n}$ and searches the time interval leading to the minimum cross-correlation between $\rho_{i,k,n}$ and its delayed version.

	% If $f_{D,l}T_s = 1/K$, $e^{j2\pi f_{D,l}kT_s} $ is a complex single-tone signal with a full cycle in $k=0,1,\cdots,K-1$. Moreover, $\rho_{i,k,n}$ is a scaled and translated version of the same single tone signal. 
	
	\begin{figure}[b]
		\centering  
		%\parbox{0.45\linewidth}{
			\includegraphics[width=0.8\linewidth]{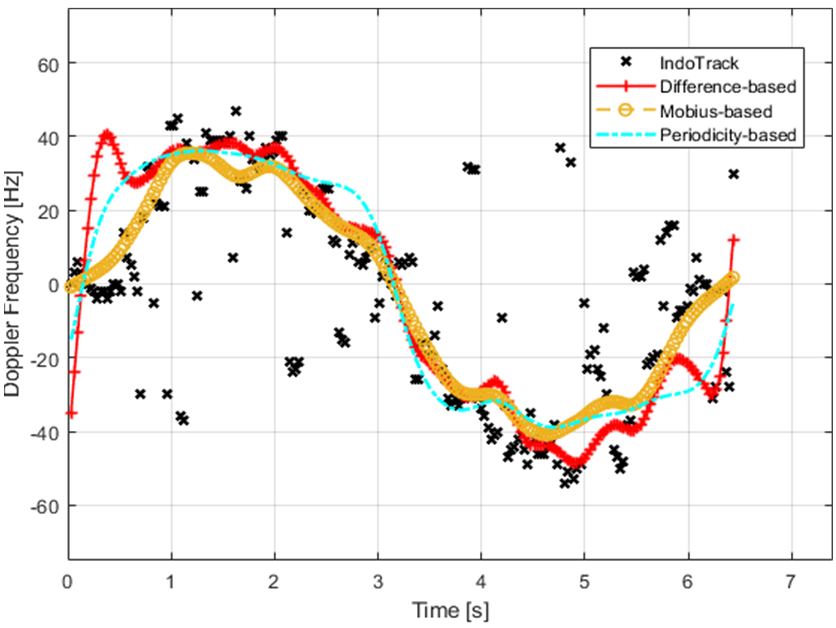}
			% \caption{Illustrating time-varying phases of \ac{lte}'s CSIs, as extracted by the NI \ac{lte} massive MIMO testbed.}
			% \label{fig: CSI phase \ac{lte}}
		%}
		% \hfill
		%\parbox{0.45\linewidth}{
			%\textcolor{white}{\ac{casr}-based Doppler estimation, in comparison with \ac{cacc}-based estimation in a classroom setting based on the Widar dataset.a classroom setting based on the Widar dataset.}  
			\captionof{figure}{\ac{casr}-based Doppler estimation, in comparison with \ac{cacc}-based estimation in a classroom setting \cite{CASR_xinyu}, based on the Widar dataset \cite{Widar2p0}. \bb{As shown in \cite{Kai_wifisensing_ICC}, one can use the ground truth locations, along with timestamps, published in \cite{Widar2p0} for calculating the true Doppler frequencies.}
				\label{fig: CASR xinyu}}  
		%}
		% \parbox{0.3\linewidth}{
			% 	\includegraphics[width=\linewidth]{figures/uts/fig_CASR_corridor.png}
			% 	% \caption{Illustrating almost random phases of Wi-Fi's CSIs, as extracted by a \ac{cots} Wi-Fi platform.}
			% 	% \label{fig: CSI phase Wi-Fi}
			% }
		% \hfill
		% \parbox{0.3\linewidth}{
			% 	\includegraphics[width=\linewidth]{figures/uts/fig_CASR_classroom.png}
			
			% 	% \label{fig: CSI phase WiFi}
			% }

		% office (left), corridor (middle), and  (right) scenarios. Figure source:}
	
	% \vspace{-1cm}
	
\end{figure}

\begin{figure*}[t!]
	\begin{center}   
		\centering
		\includegraphics[width=\textwidth]{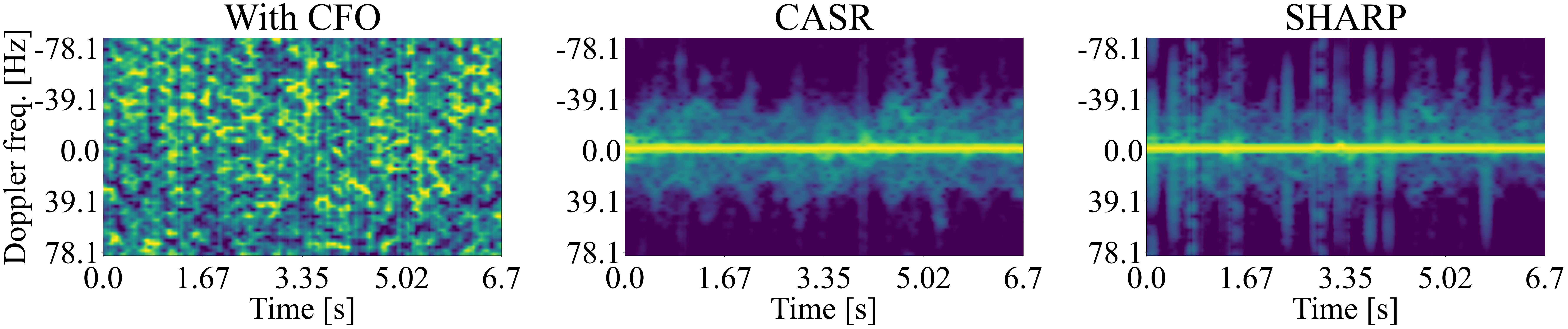}
		\caption{Example human micro-Doppler with 5 GHz carrier frequency (IEEE~802.11ac). In the leftmost plot, the spectrum is affected by \ac{cfo}. In the other plots, the phase offset was removed using \ac{casr}~\cite{CASR_xinyu} (middle) and SHARP~\cite{meneghello2022sharp} (right). Powers are normalized between 0 (blue) and 1 (yellow).}
		\label{fig: qualitative-md-5GHz}
	\end{center}
	% \vspace{-0.5cm}
\end{figure*}

When tested on the public Widar2.0 Wi-Fi dataset~\cite{Widar2p0}, the three Doppler estimation methods proposed in~\cite{CASR_xinyu} outperform the IndoTrack based on \ac{cacc}~\cite{CACC} in indoor scenarios, as illustrated in Fig.~\ref{fig: CASR xinyu}.
\bb{The results in the figure are obtained based on the open dataset Widar \cite{Widar2p0}. The data used here is collected in a classroom with a person walking in a square track and fixed communication transmitter and receiver. Thus, the Doppler frequency is varying over time. We note that each estimation in the figure is obtained based on a limited number of consecutive symbols, where the sensing parameters are approximately constant. Thus, the signal model in \eq{eq: Hnk channel matrix} applies to each estimation.} 
The main issue with \ac{cacc} is the mirrored dynamic paths, which, as seen from Fig.~\ref{fig: CASR xinyu}, can lead to false yet symmetric moving path detection. The \ac{casr} can mitigate the issue, particularly in single-dynamic-path scenarios.   

Recently, the performance bounds for Doppler sensing using the CASR method are provided in \cite{hu2024performance}, via deriving closed-form CRLB expressions. The impacts of numerous system parameters on Doppler sensing are disclosed. Waveform optimization based on the insights is also conducted, achieving substantial performance improvement.   

To utilize the \ac{csi} ratio for multi-target sensing, a Taylor series-based method is proposed in \cite{CASR_zhitongTaylor}. A general model with multiple dynamic targets is considered in the work. The second-order complex Taylor series of the \ac{csi} ratio is derived with a closed-form expression in terms of the vector of sensing parameters to be estimated. This reveals the unique patterns embedded in the differences between \ac{csi} ratios over different snapshots, facilitating super-resolution
multi-target Doppler sensing. 
\bbnote{In \cite{CASR_multiSense}, instead, \ac{ica} is used to separate the multiple targets contribution to the \ac{cfr}, thus extending \ac{casr} to multitarget scenarios but entailing higher computational cost.}

\bb{We remark that when more than one transmitting antennas are used, virtual antenna channels can be created at the sensing receiver side as in conventional MIMO radar processing. Referring to the channel matrix in \eq{eq: Hnk channel matrix}, one can see that vectorizing the matrix can generate a new steering vector, which can be seen as the Kronecker product of the original steering 
	vectors. Each entry of the vectorized channel vector can be seen as the \ac{csi} for a virtual channel.  	
	Nevertheless, we note that this spatial-domain processing does not change the fact that the nuisance caused by the random offsets, i..e, $\xi_{i,k}$ in \eq{eq: Hnk channel matrix}, is still common to all entries in the \ac{csi}. 
	This enables us to continue applying the offset elimination methods reviewed in this section to suppress the random offsets caused by clock asynchronism. However, sensing methods based on virtual channels after canceling offsets may need further investigation.}

% \bbnote{MultiSense \cite{CASR_multiSense},} a linear combination of signals over transceiving antennas is introduced to identify a reference signal that is approximately independent of dynamic paths. 
% If we stack $h_{ikn}$ in (\ref{eq: h_ikn}) over $n$ and assume that the first $L'$ paths are dynamic, then we obtain something looking like 
% \begin{align}
	% 	\mathbf{h} = offset(\mathbf{A}\mathbf{b} + \mathbf{h}_{\mathrm{s}}), 
	% \end{align}
% If we vectorize the channel matrix in (\ref{eq: Hnk channel matrix}), it can be seen as a linear combination of steering vector matrix with 

\subsection{Offsets Cancellation Methods: Delay-domain Processing Techniques}\label{sec:del-dom}

Delay-domain techniques remove \ac{to} and \ac{cfo} by exploiting the fact that \textit{they do not change in different signal propagation paths}, since they depend on the receiver hardware. 
Such techniques are, therefore, independent of the number of antennas at the transmitter or receiver devices, and they can work in \ac{siso} systems (where cross-antenna techniques are not applicable) or systems equipped with a single phased array.
To see this, we simplify the \ac{cfr} model for an \ac{ofdm} system given in Eq.~(\ref{eq: Hnk channel matrix random phase}) using $M=1$ and $N=1$.  
The channel gain for the $k$-th symbol at subcarrier $i$ is written as
\begin{equation}\label{eq: siso-cfr}
	H_{i,k} = \xi_{i,k} \sum_{l=1}^L b_l e^{-j2\pi i \Delta_f \tau_l} e^{j2\pi f_{D, l} k T_s}.
\end{equation}
The above model also holds if the transmitter and receiver are equipped with phased arrays of multiple antennas connected to the same RF chain. In this case, we consider the overall beamforming gain to be included in the coefficients $b_l$.
% due to the transmitter and receiver beamforming vectors and the channel beam steering vectors to be included in the coefficients $b_l$.
Eq.~(\ref{eq: siso-cfr}) suggests that a viable method to compensate for the phase offsets is to use one of the propagation paths as a reference: The phase differences concerning the reference path only depend on the propagation path delay and Doppler shift since $\xi_{i,k}$ cancels out. \rev{Hence, provided that the delay and Doppler of the reference path are known (e.g., because it is static so that the Doppler shift is absent and its delay can be estimated once using localization), estimation of the sensing parameters can be performed with low complexity algorithms.} This idea is similar to the one behind spatial domain (i.e., cross-antenna) processing techniques discussed in \secref{sec: spatial-domain}, but it does not require the use of multiple antennas, leveraging multipath resolution in the delay domain instead.

% Note that computing phase differences with respect to a reference path will result in obtaining an offset-free version of the \ac{cfr}, but leads to the estimation of \textit{relative} delay and Doppler parameters. 
The \ac{cfr} contribution of the reference path, $\bar{H}_{i,k}$, is expressed as $\bar{H}_{i,k} = \xi_{i,k} b_1 e^{-j2\pi i \Delta_f \tau_1  }e^{j2\pi f_{D, 1} kT_s}$,
% \begin{equation}\label{eq: refpath}
	%     \bar{H}_{ik} = \xi_{i,k} b_1 e^{-j2\pi i \Delta_f \tau_1  }e^{j2\pi f_{D, 1} kT_s},
	% \end{equation}
where we assume by convention that the reference path has index $l\!=\!1$. The offset-free \ac{cfr} $ \hat{H}_{i,k}$ is obtained by multiplying the \ac{cfr} in Eq.~(\ref{eq: siso-cfr}) by the complex-conjugate (to obtain phase differences) of the reference path's \ac{cfr}, i.e., 
\begin{equation}\label{eq: h-offfree}
	\hat{H}_{i,k} = H_{i,k} \bar{H}^*_{i,k}= b_1\sum_{l=1}^{L} b_l  e^{-j2\pi i \Delta_f (\tau_l - \tau_1)  }e^{j2\pi (f_{D, l} -f_{D, 1} )  kT_s}.
\end{equation}
In Eq.~(\ref{eq: h-offfree}), the delay and Doppler parameters of path $l$ have been replaced by the relative quantities $\tau_l \!-\! \tau_1$ and $f_{D, l} \!-\! f_{D, 1}$. \rev{However, unlike in spatial-domain techniques (see \secref{sec: spatial-domain}) here, this does not increase the subsequent parameter estimation complexity for the following two reasons. First, both in \ac{los} and \ac{nlos} conditions, relative delays $\tau_l \!-\! \tau_1 $ can be easily used to track targets by exploiting the well-known bistatic radar geometry if the distance between Tx and Rx is known (see, e.g., \cite{samczynski20215g}). Second, if the reference path used to compensate for \ac{cfo} is \textit{static}, we have $f_{D, 1} \!=\! 0$, which means that the relative Doppler shifts in~\eq{eq: h-offfree} become equal to the absolute ones. This allows for the direct estimation of the Doppler frequency, avoiding the ambiguity of relative frequency shift.}

% In the following, we present the delay-domain processing techniques proposed in \cite{meneghello2022sharp, pegoraro2023jump}. To simplify the description, we start by assuming the presence of a \ac{los} propagation path and show how it can be used to compensate for phase offsets. Then, we relax this assumption by showing that similar processing can be applied when a \ac{los} path is not available, as long as other static reflections are present. 

% When the \ac{los} path is available, it is natural to use it as a reference since it is typically stronger than other paths, thus providing a less noisy compensation of phase offsets. Moreover, with static transmitter and receiver, the \ac{los} is not affected by Doppler shift. However, delay-domain techniques work with any static path 
In the following, we detail SHARP, the delay-domain phase offsets compensation technique used in~\cite{meneghello2022sharp}, for \ac{ofdm} IEEE~802.11ac/ax systems operating on the $5$~GHz Wi-Fi bands. Delay-domain processing is also used in \cite{pegoraro2023jump}, which includes a hybrid approach and will be presented in \secref{sec: hybrid-methods}.
% , and JUMP, the algorithm presented in \cite{pegoraro2023jump}, for an \ac{sc} system based on IEEE~802.11ay ($60$~GHz carrier). 

% In~\cite{meneghello2022sharp} a custom processing technique has been proposed to compensate for the phase offsets by working on each channel sample independently and leveraging the multi-path effect. The approach has been developed and evaluated on IEEE 802.11ac (Wi-Fi 5) networks. The main intuition is that the different components of the multi-path propagation, i.e., the different copies of the transmitted signal collected at the receiver, are affected by the same phase offsets as they are processed simultaneously by the same hardware at the receiver. Hence, being able to separate the multi-path components, one can use one of them as a reference for the other components, thus mitigating the impact of the impairments. In particular, the strongest path is considered as a reference in~\cite{meneghello2022sharp}. The assumption under the application of this processing to the sequence of channel snapshots is that the strongest path remains almost unchanged over the acquisition time: This allows maintaining the relationship among the different components of the multi-path in time.

% An example of offset cancellation using a reference path is SHARP, an \ac{ofdm} sub-$7$~GHz \ac{isac} system~\cite{meneghello2022sharp}. 
In SHARP, the strongest propagation path is used as the reference by reasonably assuming that it corresponds to the \ac{los} path and remains almost unchanged over a short acquisition time. % and, in turn, the processing does not alter the relationships between each multi-path component in subsequent channel estimates. 
The main challenge of applying delay-domain phase offset removal in sub-$7$~GHz systems is their low multipath resolution due to the small transmission bandwidth (typically $40$ to $160$~MHz). This makes it hard to separate the multiple propagation paths by simply applying inverse Fourier transformations over the time dimension. To address the challenge, \cite{meneghello2022sharp} proposed a custom multipath resolution technique based on the compressive sensing-based optimization problem
\begin{equation}\label{eq:minimization_problem}
	\mathbf{r}_{k} = \argmin_{\tilde{\mathbf{r}}} ~ \left| \left| \mathbf{H}_{k} - \mathbf{T}  \tilde{\mathbf{r}}\right| \right|_2^2 + \lambda \left| \left| \tilde{\mathbf{r}} \right| \right| _1,
\end{equation}
where $\mathbf{H}_{k} = [H_{0,k}, \dots, H_{S-1,k}]^{\mathrm{T}}$ and $\mathbf{T}$ is an ($S\times L'$)-dimensional dictionary matrix with element $(i, l)$ set to $T_{i,l} = e^{-j2\pi i \tau_{l, {\rm tot}}/T_O}$ and $\tau_{l, {\rm tot}}$ with $l=0, \dots, L'-1$ is a set of \textit{candidate} combined timing offsets, including the propagation delay, $\tau_l$ and the \ac{to}, $\TO{k}$. $L' > L$ is the parameter that defines the grid of possible multi-path components included in the matrix $\mathbf{T}$.
The vector $\mathbf{r}_k$ (and, in turn, $\mathbf{\tilde{r}}$) is $L'$-dimensional and represents the subcarrier-independent terms of the offset, including the \ac{cfo} and the random phase offset, i.e., element $l$ of $\mathbf{r}_k$ is $r_{k,l}=\beta_k e^{j2\pi \FO{k} kT_s} b_l$. Note that in our model in Eq.~(\ref{eq: Hnk channel matrix random phase}), $b_l$ includes a carrier phase term, $e^{-j2\pi f_c\tau_l}$ with $f_c$ being the carrier frequency, which is instead made explicit in~\cite{meneghello2022sharp}.
The position of the non-zero entries in vector $\mathbf{r}_k$, obtained by solving Eq.~\ref{eq:minimization_problem}, indicate the estimated multi-path components out of the $L'$ ones. The estimate of the multi-path decomposition of matrix $\mathbf{H}_k$, for each \ac{ofdm} subcarrier $i$, is obtained by combining $\mathbf{r}_k$ and the $i$-th row of $\mathbf{T}$, denoted by $\textbf{T}_i$, through the Hadamard product, i.e., $\textbf{X}_{i,k} = \textbf{T}_i^{\mathrm{T}} \circ \mathbf{r}_k$. Finally, $\hat{H}_{i,k}$ in \eq{eq: h-offfree} is obtained as $\hat{H}_{i,k} = \sum_{l=1}^{L'}\mathbf{X}_{i,k}X_{i,k,0}^*$, where $X_{i,k,0}$ is the first element of vector $\mathbf{X}_{i,k}$ and is associated with the strongest multipath component.

Example results on human micro-Doppler spectrogram extraction, comparing \ac{casr} and SHARP, are shown in Fig.~\ref{fig: qualitative-md-5GHz}. The results refer to an IEEE~802.11ac testbed with a $5$~GHz carrier frequency. 

% Following the procedure in~\cite{meneghello2022sharp}, the reconstructed \ac{cfr} is expressed as
% \begin{equation}\label{eq:hest}
	% 	\hat{H}_{n}[k] = A_{p^*}(k)\sum_{p'=0}^{P'-1} A_{p'}(k) e^{-j2\pi (f_c + n/T)(\tau_{p'}(k)-\tau_{p^*}(k))},
	% \end{equation} 
% where $A_{p^*}(k)$ and $\tau_{p^*}(k)$ are the attenuation and delay of the strongest path $p^*$ in the decomposition. $\hat{H}_{n}[k]$ of \eq{eq:hest} represents the \ac{cfr} estimate for subcarrier $n$, where the phase offset has been removed and the contribution of each path is modulated according to the amplitude and phase of path $p^*$. 

% The above method could be improved upon by accounting for the fractional part of $\tau_k^r$, e.g., by upsampling the \ac{cir} magnitude profile before correlation. 
% However, this would also incur an additional processing complexity. Another improvement proposed in \cite{pegoraro2023jump} consists in enhancing the \ac{to} estimation using the \ac{cir} profiles obtained with different transmitter \acp{bp}.

% Finally, we remark that although the above technique was developed in a wideband \ac{mmwave} system~\cite{pegoraro2023jump}, it can be also applied to sub-$7$~GHz bands. To avoid the lower delay resolution in such cases, one may apply higher resolution estimation techniques to obtain the \ac{cir} magnitude, e.g., using \ac{music} on the \ac{cfr}.  

\subsection{Offsets Estimation Methods}

% \bbnote{These methods utilize the strong \ac{los} path to estimate asynchronism factors. 
	% \cite{Kai_wifisensing_ICC}:
	% \cite{Uplink_jinbo_TVT}
	% \cite{Uplink_xuchen2023kalman}
	% }

Different from the above methods using the spatial or time domain features to cancel the random phase offsets caused by the clock asynchronism, i.e., $\xi_{i,k}$ in (\ref{eq: Hnk channel matrix random phase}), the designs in \cite{CASR_multiSense, Uplink_jinbo_TVT, Kai_wifisensing_ICC} seek to estimate and compensate the offset for high-performance Doppler sensing. The rationale can be seen from Eq.~(\ref{eq: r_ikn}) -- if only static paths are left in the denominator, the Doppler estimation would not suffer from the non-linearity caused by the \ac{casr} anymore. 

A spatial-domain linear combination is adopted in~\cite{CASR_multiSense, Uplink_jinbo_TVT} to suppress the signals associated with the dynamic paths. The starting point of such designs is the transceiving relationship in the spatial domain, as embedded in the steering vectors in Eq.~(\ref{eq: Hnk channel matrix random phase}). 
\bb{More specifically, $\mathbf{H}_{i,k}$ therein can be seen as a linear combination of $\mathbf{a}(N,\phi_{l})\mathbf{a}^{\mathrm{T}}(M,\theta_{l})$ which is the outer product of two steering vectors concerning transmitting and receiving arrays. This outer product mainly conveys the spatial information of a path. Based on the matrix manipulation rule that $\mathrm{vec}(\mathbf{ABC}) = \mathbf{C}^{\mathrm{T}}\otimes\mathbf{A}\mathrm{vec}(\mathbf{B})$, where $\otimes$ denotes the Kronecker product, we can vectorize $ \mathbf{H}_{i,k}$ and obtain  
%\begin{strip}
    \begin{align}
		&\mathrm{vec}(\mathbf{H}_{i,k})\notag\\
  &= \underbrace{ \left[ \mathbf{a}(M,\theta_{1})\otimes \mathbf{a}(N,\phi_{1}),\cdots,\mathbf{a}(M,\theta_{L})\otimes \mathbf{a}(N,\phi_{L}) \right] }_{\mathbf{A}} \mathbf{b}_{i,k},
	\end{align}
 %\end{strip}
	where $\mathbf{b}_{i,k}$ is a column vector collecting the coefficients of the $L$ paths in \eq{eq: Hnk channel matrix random phase}). The above conversion is similar to conventional MIMO radar processing, where $M$ Tx antennas and $N$ Rx antennas can lead to $MN$ virtual antennas. Using array signal processing theories, we see that it is theoretically feasible to identify the null space of dynamic paths. Projecting the CSI signals onto the null space can then facilitate the estimation of channel coefficients associated with the static path.}

To achieve this, an optimization is performed in~\cite{CASR_multiSense} to find the linear combination over virtual antennas that minimizes the ratio of respiration energy to the overall signal energy. 
The respiration energy is the signal energy in the frequency range caused by human respiratory movement, which can be determined based on system configurations, such as carrier frequency and the transceiver's geometric relationship. 
In \cite{Uplink_jinbo_TVT}, the MUSIC algorithm estimates the \acp{aoa} of static and dynamic paths. It is shown in the work that, in the presence of a strong \ac{los} path, the Bartlett beamformer would form a peak at the \ac{los} path. This feature is then used to differentiate the \ac{los} path's \ac{aoa} from those of dynamic paths. The null space of the dynamic paths in the spatial domain is then constructed to extract the \ac{los} path, which facilitates the estimation of the composite random phase caused by the asynchronism, $\xi_{i,k}$. Similarly, the null space method extracts each dynamic path by suppressing others, enabling the estimation of their Doppler frequencies.   

If a strong \ac{los} path exists and only the Doppler information is of interest, estimating and compensating for the asynchronism offsets can be quite effective by fully taking advantage of the \ac{los} path. This is demonstrated in \cite{Kai_wifisensing_ICC}. 
Treating the first path in Eq.~(\ref{eq: Hnk channel matrix}) as the strong \ac{los} path, then $ b_1 e^{-j 2\pi i (\tau_{1} + \tau_{o,k} ) \Delta_f } $ can be seen as a single-tone signal indexed by the sub-carrier index $i$. Estimating the frequency of this signal gives us an estimate of the composite delay with \ac{los} path delay plus timing offset. 
We refer interested readers to~\cite{Kai_wifisensing_ICC, Kai_padeFreqEst2021TVT} for the details on the estimation method. It is based on DFT interpolation, a popular low-complexity frequency estimation method with off-grid accuracy.
With the composite delay of the \ac{los} path, i.e., $\tau_{1} + \tau_{o,k}$, estimated, signals can be coherently accumulated over the \ac{los} path, and the phase of the high-\ac{snr} accumulated signal can be used as the estimation of all other random phase offsets. This method applies to any antenna and symbol as seen from Eq.~(\ref{eq: Hnk channel matrix}).

\begin{figure*}[t!]
	\begin{center}   
		\centering
		\includegraphics[width=0.6\textwidth]{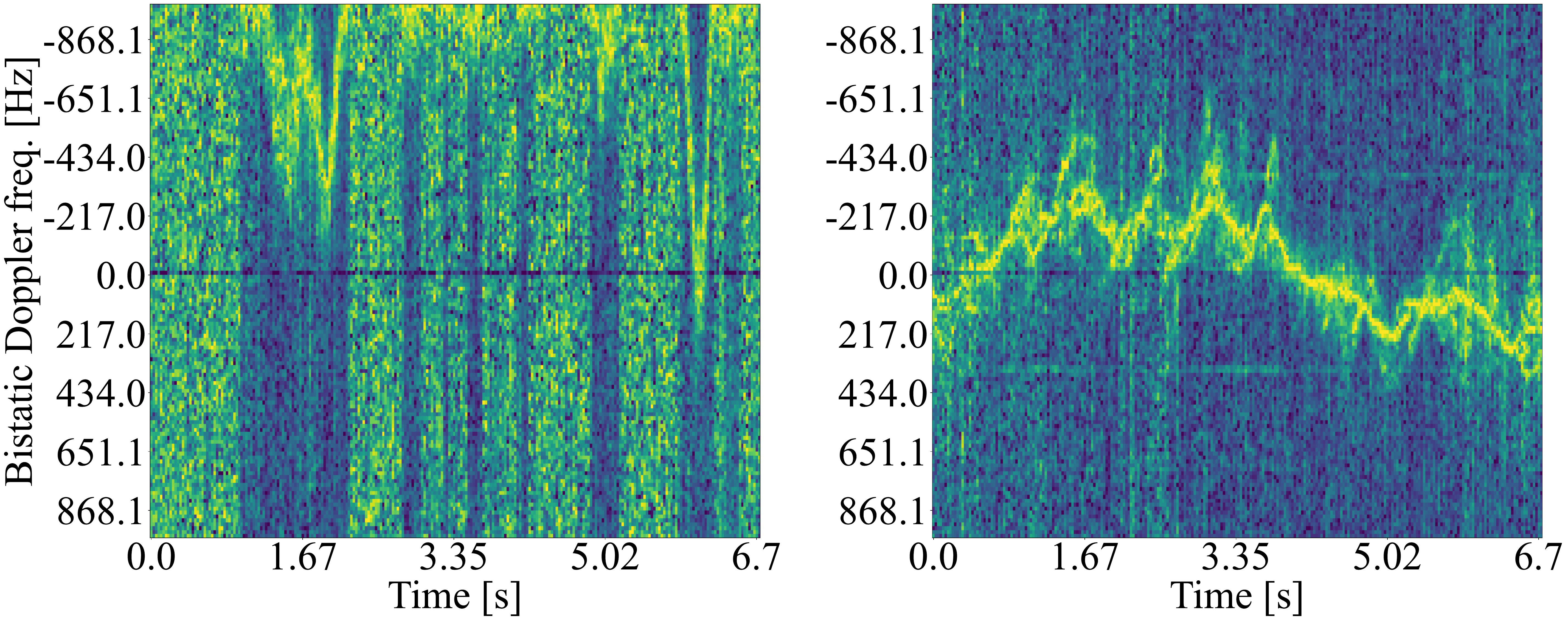}
		\caption{Example human micro-Doppler with 60 GHz carrier frequency (IEEE~802.11ay). On the left, the micro-Doppler spectrum is affected by \ac{cfo}. On the right, we apply JUMP~\cite{pegoraro2023jump} to remove it. Powers are normalized between 0 (blue) and 1 (yellow).}
		\label{fig: qualitative-md-60GHz}
	\end{center}
	\vspace{-0.5cm}
\end{figure*}

\subsection{Hybrid Methods}\label{sec: hybrid-methods}
% \rev{[19] and [36]}

In \cite{pegoraro2023jump}, a hybrid method, named JUMP, that combines \ac{to} estimation and suppression with \ac{cfo} estimation was proposed for a \ac{mmwave} system based on IEEE~802.11ay. \ac{mmwave} systems offer a wide transmission bandwidth that, in turn, leads to higher multipath resolution than in sub-$7$~GHz ones. 
% As an example, IEEE 802.11ay is based on \ac{sc}, $B=1.76$~GHz-wide channels, which lead to a delay resolution of $1/B \approx 0.57$~ns. 
This makes \ac{mmwave} systems the ideal application scenario for delay-domain techniques (see \secref{sec:del-dom}) since multipath separation is easily achieved by detecting the peaks of the \ac{cir}. 
The latter is obtained as part of the communication protocol to perform channel equalization in \ac{sc} systems (like IEEE~802.11ay), while it can be obtained via \ac{idft} in \ac{ofdm} systems. 
\rev{Recalling that $\xi_{i, k} = \beta_ke^{-j2\pi i\Delta_f \TO{k}} 
	e^{j2\pi \FO{k}kT_s}$, the \ac{cir} model corresponding to the \ac{cfr} in \eq{eq: siso-cfr} is
	\begin{equation}\label{eq:cir-model}
		h_{n,k} = \beta_{k}\sum_{l=1}^{L} b_l e^{j2\pi (f_{D,l} + \FO{k}) kT_s} \delta_{n - \tau_l - \TO{k}},
\end{equation}}
where $n$ is the index of delay-domain samples spaced by $1/B$ and $\delta_n$ is the Kronecker delta. \rev{Note that, in \eq{eq:cir-model}, using Kronecker delta functions is an approximation of the real \ac{cir} that neglects the impact of the non-ideal autocorrelation of pilot sequences and sampling points in \ac{sc} systems, or the finite bandwidth in \ac{ofdm} ones. We neglect such non-idealities as done in~\cite{pegoraro2023jump} to maintain the same notation.} 

From \eq{eq:cir-model}, one can see that the effect of \ac{to} is a \textit{shift} of the \ac{cir} by $\TO{k}$, while the \ac{cfo} affects the phase of the \ac{cir}. \cite{pegoraro2023jump} presents a hybrid algorithm that first estimates and compensates for the \ac{to}, then cancels out the \ac{cfo} using a delay-domain approach. The two processing steps are detailed next.
\begin{enumerate}
	\item \textit{\ac{to} estimation and compensation:} The multiple propagation paths appear as peaks in the magnitude of the estimated \ac{cir} at the receiver, which can be detected, e.g., by defining a suitable threshold or by more advanced adaptive methods~\cite{pegoraro2023jump}. Then, the first detected path, having the lowest delay, is the \ac{los} path to be used as a reference. The \ac{to} appears as a common shift of all propagation paths, so it can be removed by estimating the \ac{los} path peak location, equal to $\tau_1 + \TO{k}$, and applying an opposite shift to the \ac{cir} by convolving it with~$\delta_{n + \tau_1 + \TO{k}}$. 
	\item \textit{\ac{cfo} cancellation:} This second step is based on the delay-domain approach. The phase of a static reference path, i.e., having $f_{D, 1} = 0$, is extracted. \rev{Such phase contains the nuisance due to $f_{o, k}$, and $\angle{\beta_k}$,} so the \ac{cfo} and random phase offset are removed from the \ac{cir} by computing $\hat{h}_{n,k} = \tilde{h}_{n,k} e^{-j\angle \tilde{h}_{1,k}}$, where $\tilde{h}_{1,k}$ is the \ac{to}-free \ac{cir} resulting from point 1).  
\end{enumerate}
The resulting \ac{cir} allows estimating relative delays and Doppler frequencies with low complexity algorithms. As an example, in \cite{pegoraro2023jump}, this is done by applying peak detection to the \ac{cir} magnitude to obtain $\tau_l - \tau_1$ with $ l=1, \dots, L$ from the locations of the peaks. Then, $f_{D, l}$ are estimated by performing a \ac{dft} over a window of $K$ \ac{cir} samples. The Doppler frequencies correspond to the peaks in the Doppler spectrum, computed as the squared magnitude of the \ac{dft}.

Example results showing the effect of \ac{cfo} compensation on a human micro-Doppler spectrogram are shown in Fig.~\ref{fig: qualitative-md-60GHz}. The experimental data was obtained using a \ac{fpga}-based IEEE~802.11ay system with $60$~GHz carrier frequency.

The above offsets compensation technique is then extended to the case where the \ac{los} path is not always available, e.g., because it is temporarily blocked by an obstacle. 
The approach relies on the assumption that the channel contains multipath reflections on static objects, which are slowly time-varying compared to the packet transmission rate, which is a common assumption for radar and \ac{isac}. The technique leverages the similarity of the \ac{cir} magnitude profiles across subsequent packets and performs the following steps
\begin{enumerate}
	\item Estimation of the \textit{relative} \ac{to} between subsequent \ac{cir} estimates, defined as $\tau_{o}^r \triangleq \TO{k} - \TO{k-1}$. This can be done, e.g., by locating the correlation peak for different candidate shift values between the \ac{cir} amplitude profiles $|h_{n,k}|$ and $|h_{n,k-1}|$.
	\item Compensation of the relative \ac{to} by shifting $h_{n,k}$ by $\tau_{o}^r$. When applied sequentially to subsequent \ac{cir} estimates for $k=0, \dots, K-1$, this step incrementally removes the \ac{to}. 
\end{enumerate}
Thanks to 1), any static path can be used as a reference since the \ac{cir} samples across time share the same timing reference.
\rev{Selecting a reference path different from the LOS may be challenging when \ac{cfo} is present since the distinction between static paths and dynamic ones cannot be made based on the corrupted Doppler shift. In~\cite{pegoraro2023jump}, this is solved by applying a tracking algorithm to the multipath components of the \ac{cir} to estimate the location of the scatterer in the 2D Cartesian plane after \ac{to} has been compensated for. Static paths are identified as the ones whose position does not change significantly across time, and the reference path is selected as the static path with the highest amplitude.}

To initialize the algorithm, \cite{pegoraro2023jump} assumes that the sensing operation starts in a \ac{los} condition, where the \ac{to} can be removed, or that the \ac{los} becomes available at least once during the system operation. \rev{Under this assumption, the compensation of the \textit{relative} \acp{to} between subsequent packets by steps 1)-2) allows maintaining the \ac{los}-based timing reference even when transitioning to \ac{nlos} conditions for intermittent time intervals.}
Note that the above technique depends on the fact that transitions between \ac{los} and \ac{nlos} conditions are slow compared to variations of the  \ac{cir} magnitude profile. This makes it appealing for indoor scenarios, in which it is likely to have multiple static propagation paths (e.g., from reflections on the walls). Conversely, due to its sensitivity to abrupt changes in the multipath profile, this technique may not be suitable for highly dynamic outdoor scenarios. 

\newcolumntype{Y}{>{\centering\arraybackslash}X}

\begin{table*}[!t]
	\footnotesize
	\centering
	\caption{Comparison of different bi-static sensing Techniques (In the Processing Domain column, S, D, and F stand for ``Spatial," ``Delay," and ``Frequency," respectively).}
	\label{tab: compare all methods}
	\begin{tabularx}{\textwidth}{c|c|Y|Y|Y|Y|Y|Y}
		\hline
		\textbf{Type} 
		& \textbf{Methods} 
		& \textbf{Strong \ac{los}}
		& \textbf{Multiple Targets} 
		& \textbf{Single Antenna}
		& \textbf{Processing Domain}
		& \textbf{Doppler-\ac{aoa} Only}
		& \textbf{OTA}
		\\        
		\hline
		
		\multirow{6}{*}{\textbf{I(S)}} 
		& Indotrack \cite{CACC} 
		& \checkmark 
		& \xmark
		& \xmark 
		& S 
		& \checkmark 
		& \checkmark
		\\
		\cline{2-8}

		& WiDFS \cite{CACC_zhongqin} 
		& \checkmark 
		& \xmark
		& \xmark
		& S
		& \checkmark 
		& \checkmark 
		\\
		\cline{2-8}

		& \ac{cacc} \cite{CACC_ni2020uplink} 
		& \checkmark
		& \checkmark
		& \xmark
		& S
		& \checkmark
		& \xmark
		\\
		\cline{2-8}

		& FarSense \cite{CASR_farSense} 
		& \xmark
		& \xmark
		& \xmark
		& S
		& \checkmark
		& \checkmark
		\\
		\cline{2-8}

		& \ac{casr} \cite{CASR_xinyu} 
		& \xmark
		& \xmark
		& \xmark
		& S
		& \checkmark
		& Widar %\cite{Widar2p0}
		\\
		\cline{2-8}

		& \ac{casr} \cite{CASR_zhitongTaylor} 
		& \xmark
		& \checkmark
		& \xmark
		& S
		& \checkmark
		& WiDFS %\cite{CACC_zhongqin}
		\\
		\hline
		
		\textbf{I(D)} 
		& SHARP \cite{meneghello2022sharp} 
		& \xmark
		& \xmark
		& \checkmark
		& D
		& \xmark
		& \checkmark \cite{meneghello2023CSI}
		\\
		\hline
		
		\multirow{3}{*}{\textbf{II}} 
		& multiSense \cite{CASR_multiSense} 
		& \xmark
		& \checkmark
		& \xmark
		& S, F
		& \checkmark
		& \checkmark
		\\
		\cline{2-8}

		& \cite{Uplink_jinbo_TVT} 
		&  \checkmark 
		& \checkmark
		& \xmark
		& S
		& \checkmark
		& \checkmark
		\\
		\cline{2-8}

		& \cite{ Kai_wifisensing_ICC} 
		& \checkmark
		& \checkmark
		& \checkmark
		& F
		& \checkmark
		& Widar %\cite{Widar2p0}
		\\
		\hline
		
		\textbf{III} 
		& JUMP \cite{pegoraro2023jump} 
		& \xmark
		& \checkmark
		& \checkmark
		& D
		& \xmark
		& \checkmark
		\\
		\hline
	\end{tabularx}
\end{table*}

\subsection{Comparisons and Remarks}

Table~\ref{tab: compare all methods}
compares all methods reviewed above regarding their requirements and features, where OTA is short for ``over the air". The three types are stated at the beginning of Section \ref{sec: review bi-static sensing techniques}, where the first type is further divided into two sub-types, with the spatial-domain methods, under the type I(S) in the table reviewed in Section \ref{sec: spatial-domain}, and the delay-domain method, under I(D), reviewed in Section \ref{sec:del-dom}. In Table~\ref{tab: compare all methods}, the three domains are mainly referred to as where the offsets are addressed. 
As mentioned earlier, most works are predominantly performed in the spatial domain. However, some recent works incorporate other domains to relieve the reliance on multiple antennas and explore the degrees of freedom available. 

Moreover, not all methods can sense multiple targets concurrently moving in the environment. Specifically, \ac{cacc} and delay-domain techniques support multiple targets, subject to having enough bandwidth and/or \ac{aoa} resolution to resolve the corresponding multipath components. Conversely, \ac{casr} is mainly applicable to single-target scenarios and requires special processing to separate the multiple contributions, as shown in~\cite{CASR_zhitongTaylor, CASR_multiSense}.

As a sensing-related application, positioning or localization in 5G and beyond networks is emerging as a hot research topic \cite{palacios2022low,nazari2023mmwave}. In such applications, timing offset is of particular interest, as the time of arrival is typically a key sensing parameter used in localization algorithms. In \cite{palacios2022low}, a time difference of arrival is proposed to counteract timing offsets. In \cite{nazari2023mmwave}, a novel maximum likelihood estimator is developed to estimate timing offset and other sensing parameters, such as multi-path angles. While the timing offset can be estimated using these approaches, they do not sufficiently treat the frequency offset and random phase. In bi-static sensing, the methods reviewed in Sections III-B to III-D may be further applied on top of these approaches \cite{palacios2022low,nazari2023mmwave}; alternatively, it can also be interesting to extend the methods \cite{palacios2022low,nazari2023mmwave} with the other two types of offsets taken into account.

\bb{We remark that the ideas behind the methods reviewed in this section are promising to be applicable in the 6G context for addressing clock offsets, although quite a number of them are developed in the Wi-Fi sensing background. 
	This is because, regardless of network differences, the MIMO-OFDM waveform is used by both cellular and Wi-Fi networks, and all the reviewed methods can apply to a general MIMO-OFDM setup. 
	However, it is noteworthy that, when applying these methods to the 6G network,
	adaptions/changes may be necessary due to the potentially different features of 6G networks compared with Wi-Fi, as highlighted below. 	
	\begin{itemize}
		\item The 6G base station may be equipped with the so-called massive MIMO arrays, which can be efficiently applied to achieve great spatial resolution and separate multiple objects to be sensed. Millimeter wave systems will likely be hybrid arrays with analog beamforming before their digital counterparts. This means that the cross-antenna techniques \cite{CACC,CACC_zhongqin,CACC_ni2020uplink,CASR_farSense,CASR_xinyu,CASR_zhitongTaylor} will need to be extended to deal with \textit{cross-beam} signals. Moreover, analog beamforming may be designed to facilitate better solutions to addressing clock offsets. 
		
		\item  6G is expected to have a much larger frequency bandwidth than Wi-Fi, leading to more degrees of freedom in the time-frequency-domain resources. These can further improve some of the methods reviewed in this section. For example, the large bandwidth can be enjoyed by SHARP \cite{meneghello2022sharp} to facilitate much higher resolution in the range domain, hence addressing clock offset issues more effectively. As another example, the waveform design in the time-frequency domain can be resorted to, as performed in \cite{hu2024performance}, for better offset estimation performance. 
		
		\item  6G networks may be deployed in both indoor and outdoor. For indoor applications, the channels will be similar to WiFi channels, and techniques validated on a WiFi platform can be expected to work effectively on 6G networks. The 6G outdoor channels can be substantially different from Wi-Fi ones. This makes channel-dependent methods, such as  \cite{CASR_multiSense} in indoor Wi-Fi scenarios, not directly applicable to 6G. However, the idea may be borrowed. For example, instead of using the respiratory movement for estimating the optimal combination over antennas (see Section III-C for more details), the varying channels caused by a car's movement in a rural area can be relied on in 6G outdoor sensing to achieve similar goals.     
	\end{itemize}
}

\section{Future research directions and open problems}\label{sec:future-work}

We have seen promising advances in addressing the clock asynchronism in bi-static \ac{isac}. However, admittedly, many open problems still need more attention from the academia and industry communities. 

\begin{figure}[!t]
	\centering
	\includegraphics[width=\linewidth]{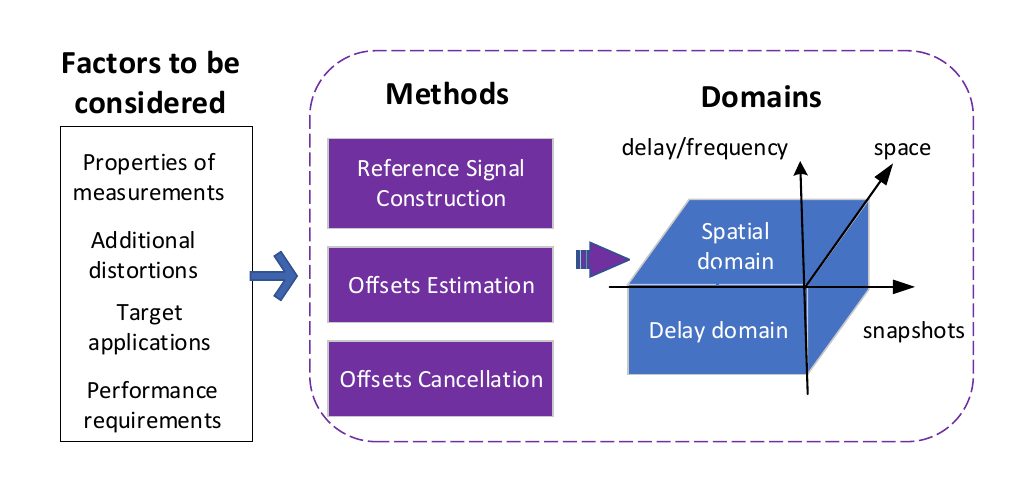}
	\caption{A generalized framework for solutions to clock asynchronism in bi-static sensing. }
	\label{fig: general}
	
	% \vspace{-1cm}
\end{figure}

The discussion on the techniques above enables us to formulate a comprehensive framework, laying the groundwork for future research at an elevated level. This framework, depicted in Fig.~\ref{fig: general}, is two-dimensional. One dimension encompasses signal processing methods, including reference signal construction, offset estimation, and offset cancellation via cross-correlation and signal ratio. The second dimension pertains to the domains where signal processing is primarily applied, with the typical space and delay domains. These methods have the flexibility to be implemented individually or in combination across one or more domains. While existing works, as explored in Section \ref{sec: review bi-static sensing techniques}, have substantiated feasibility and demonstrated potential, a thorough examination of the advantages and disadvantages of these methods and their application domains reveals opportunities for further refinement and exploration. A summary of the examination is as follows:
\begin{itemize}
	\item Constructing a reference signal can simplify the subsequent sensing parameter estimation, but it will inevitably introduce errors;
	\item Offset estimation enables flexible signal processing. However, it also faces estimation errors, which could be larger than those in reference signal construction;
	\item Offset cancellation exploits signal structures and removes offsets without introducing errors. However, it typically leads to more parameters to be estimated (signal cross-correlation) or a nonlinear problem (signal ratio);
	\item Spatial-domain processing well retains independence of paths in delay and Doppler domain, but multiple receiving chains are required, which may not be cost-effective, particularly for millimeter wave systems;
	\item Delay-domain processing only requires a single receive antenna, but obtaining CIR is sometimes challenging in OFDM systems when measurements are only available at part of the subcarriers. In addition, each CIR may be the superposition of multiple paths when delays are off-grid.
\end{itemize}
Therefore, a proper combination can be designed by jointly considering the respective advantages and disadvantages of these methods and domains, together with additional signal and system factors, as shown in Fig.~\ref{fig: general}. By delineating the two-dimensional landscape of signal processing methods and their application domains, novel combinations and innovations may be achieved. Within the framework, some more specific research problems are elaborated next.

\subsubsection{Construction of a reference path} The efficient and reliable construction of a reference path is an interesting open research challenge in delay-domain processing techniques to enable phase offset compensation. 
\cite{pegoraro2023jump} proposed to apply tracking algorithms to maintain an estimate of several candidate reference paths, including first-order reflections on static objects.
However, the sensing accuracy depends on the signal quality received from the reference path and degrades with low \ac{snr}.

\subsubsection{Sensing of multiple targets} To the best of our knowledge, only two works \cite{CASR_multiSense, pegoraro2023jump} have experimentally validated phase offsets removal techniques (hybrid, delay-domain and \ac{casr}, respectively) on data collected with multiple sensing targets concurrently moving in the environment. Cross-antenna techniques do not work out of the box in such cases, and techniques to separate the contribution of the different subjects are required, as detailed in \secref{sec:CASR}. Reference path-based techniques are not affected by multiple targets. However, obtaining a reliable reference path when multiple targets are moving, possibly causing occlusion to the \ac{los}, is extremely challenging and deserves further attention from the research community.

\subsubsection{Mobility of the nodes} Performing \ac{isac} with \textit{mobile} nodes is a challenging problem in the presence of phase offsets. Notably, the assumption that the transmitter and receiver are static is critical to all three phase offset removal techniques. To our knowledge, only \cite{liu2023towards} has presented a preliminary dynamic \ac{cfr} model for Wi-Fi sensing, but it does not deal with phase offsets.
Possible approaches to tackle this problem may come from the passive radar literature, where radars mounted on moving platforms are well-studied, e.g., \cite{wojaczek2018reciprocal}.

\subsubsection{Coherently combine multiple receivers} To fully exploit the potential of pervasive \ac{isac}, algorithms that combine the information from multiple, phase-synchronized receivers need to be developed. This would boost the system's sensing resolution, overcoming the limitation imposed by the bandwidth and enabling wavelength-level sensing accuracy. To enable this, algorithms that effectively remove the different phase offsets between each transmitter-receiver must be investigated to allow the coherent fusion of the data obtained at each receiver. A preliminary study on this subject can be found in~\cite{tagliaferri2023cooperative}.

\subsubsection{Joint spatial-delay-frequency-domain processing}
From Table~\ref{tab: compare all methods}, we see that existing designs generally utilize one or two but not all three domains (spatial, delay, and frequency) for addressing the clock asynchronism. This contributes to the restrictions of existing methods, e.g., having a strong \ac{los}, large bandwidth, or a single dynamic path, that are often associated with which domain we use for phase offsets compensation. For instance, many spatial and frequency domain methods require strong \ac{los}, which is unnecessary for the delay-domain methods. However, the performance of the delay domain method may be subject to the bandwidth available, while the other domains are not. 
Jointly using the three domains can allow us to exploit and combine the advantages of different approaches, promising to remove the restrictions suffered by most existing works.

\subsubsection{Open dataset for bi-static sensing research and development}
While most works reviewed in Section \ref{sec: review bi-static sensing techniques} perform OTA experiments to validate their methods, custom configurations and hardware platforms are often heterogeneous and do not follow unified standards. For this purpose, only a few publicly available open datasets, such as those mentioned in Table \ref{tab: compare all methods}, are available. Moreover, most existing works and experiments are based on Wi-Fi protocols/platforms. Other wireless systems, such as mobile networks, have been validated in a few works, such as \cite{LTE_sensing_JieXiong}. Therefore, a critical aspect for advancing bi-static sensing further is the availability of a comprehensive open dataset based on different communications protocols in various yet typical sensing scenarios that can be used to assess and compare the performance of the proposed approaches.

\section{Concluding remarks}\label{sec:conclusion}
In this survey, we have provided a first review of the clock asynchronism issue in the bi-static sensing against the highly popular \ac{isac} background. We started by establishing the MIMO-OFDM signal model, which is widely used in modern communications systems. We then modeled and explained the impact of random offsets caused by the clock asynchronism issue on bi-static sensing. Moreover, we reviewed three major solutions to the issue, as differentiated by whether the offsets are canceled, estimated and suppressed, or treated in a hybrid way. The basic signal processing principles of the reviewed techniques are highlighted, and some interesting results are demonstrated. After comparing all reviewed methods, we drew insights into potential research directions and open problems. We hope our first survey paper on this topic can foster more research activities on maturing bi-static sensing in \ac{isac}.

\bibliographystyle{IEEEtran}
\bibliography{bib_JCAS.bib}

\begin{IEEEbiographynophoto}{Kai Wu} (M'21) is a lecturer with the Global Big Data Technologies Centre (GBDTC), School of Electrical and Data Engineering (SEDE), University of Technology Sydney (UTS), Sydney, NSW 2007, Australia. (kai.wu@uts.edu.au)
	  \end{IEEEbiographynophoto}
	  \vspace{-1.3cm}
	  \begin{IEEEbiographynophoto}
	  	{Jacopo Pegoraro} (M'23) is a postdoctoral researcher and lecturer in the Department of Information Engineering (DEI) the University of Padova, Italy. (pegoraroja@dei.unipd.it)
	  \end{IEEEbiographynophoto}
	  \vspace{-1.3cm}
	  \begin{IEEEbiographynophoto}
	  	{Francesca Meneghello} (M'22) is an assistant professor with the Department of Information Engineering (DEI) at the University of Padova, Italy. (francesca.meneghello.1@unipd.it)
	  \end{IEEEbiographynophoto}
 \vspace{-1.3cm}
 	\begin{IEEEbiographynophoto}{J. Andrew Zhang} (M'04-SM'11) is a Professor in the School of Electrical and Data Engineering, University of Technology Sydney, Australia. (andrew.zhang@uts.edu.au)
	  \end{IEEEbiographynophoto}
	   \vspace{-1.3cm}
	  \begin{IEEEbiographynophoto}{Jesus O. Lacruz}
	  	is a Research Engineer at IMDEA Networks, Spain since 2017. (jesusomar.lacruz@imdea.org)
	  \end{IEEEbiographynophoto}
	  \vspace{-1.3cm}
	  \begin{IEEEbiographynophoto}{Joerg Widmer}
	  	(%M'06-SM'10-
	  	F'20) 
	  	is Research Professor and Research Director of IMDEA Networks in Madrid, Spain. (Joerg.Widmer@imdea.org)
	  \end{IEEEbiographynophoto}
	  \vspace{-1.3cm}
	  \begin{IEEEbiographynophoto}
	  	{Francesco Restuccia} (SM'21) is an Assistant Professor of Electrical and Computer Engineering at Northeastern University, United States. (f.restuccia@northeastern.edu)
	  \end{IEEEbiographynophoto}
	  \vspace{-1.3cm}
	  \begin{IEEEbiographynophoto}
	  	{Michele Rossi} (SM'13) is a full professor at the Department of Information Engineering (DEI) and the Department of Mathematics of the University of Padova. (michele.rossi@unipd.it)
	  \end{IEEEbiographynophoto}
 \vspace{-1.3cm}
 	\begin{IEEEbiographynophoto}{Xiaojing Huang} (M'99-SM'11) is currently a Professor of Information and Communications Technology with the School of Electrical and Data Engineering and the Program Leader for Mobile Sensing and Communications with the Global Big Data Technologies Center, University of Technology Sydney (UTS), Australia. (xiaojing.huang@uts.edu.au)  %
	  \end{IEEEbiographynophoto}
 	\vspace{-1.3cm}
 	\begin{IEEEbiographynophoto}{Daqing Zhang} is a Professor with Peking University, China and IP Paris, France. 
	 He is a Fellow of IEEE and Member
	 	of Academy of Europe. (daqing.zhang@telecom-sudparis.eu)
	  \end{IEEEbiographynophoto}
 	\vspace{-1.3cm}
 	\begin{IEEEbiographynophoto}{Giuseppe Caire} (S’92 – M’94 – SM’03 – F’05) is an Alexander von Humboldt Professor with the Faculty of Electrical Engineering and Computer Science at the Technical University of Berlin, Germany. (caire@tu-berlin.de)
	  \end{IEEEbiographynophoto}
 	\vspace{-1.3cm}
 	\begin{IEEEbiographynophoto}{Y. Jay Guo} (F’14) is a Distinguished Professor and the funding Director
	 	of Global Big Data Technologies Centre (GBDTC) at the
	 	University of Technology Sydney (UTS), Australia. He is the
	 	founding Technical Director of the New South Wales (NSW)
	 	Connectivity Innovation Network (CIN). (jay.guo@uts.edu.au)
	  \end{IEEEbiographynophoto}

\end{document}

%% file: ACRONYMS.tex
\acrodef{prop}[\textit{MIMORPH}]{MIMO Radio Platform for Heterogeneous wireless systems}

\acrodef{3gpp}[3GPP]{3rd Generation Partnership Project}
\acrodef{abft}[A-BFT]{Association Beamforming Training}
\acrodef{ack}[ACK]{Acknowledge}
\acrodef{adc}[ADC]{Analog-to-Digital Converter}
\acrodef{af}[AF]{Ambiguity Function}
\acrodef{dac}[DAC]{Digital-to-Analog Converter}
\acrodef{aoa}[AoA]{Angle of Arrival}
\acrodef{aod}[AoD]{Angle of Departure}
\acrodef{agc}[AGC]{Automatic Gain Control}
\acrodef{ap}[AP]{Access Point}
\acrodef{amc}[AMC]{Advanced Mezzanine Card}
\acrodef{awv}[AWV]{Antenna Wave Vector}
\acrodef{axi}[AXI]{Advanced eXtensible Interface}
\acrodef{ber}[BER]{Bit Error Rate}
\acrodef{bft}[BFT]{Beamforming Training}
\acrodef{bp}[BP]{Beam Pattern}
\acrodef{bpsk}[BPSK]{Binary Phase-Shift Keying}
\acrodef{brp}[BRP]{Beam Refinement Phase}
\acrodef{cs}[CS]{Compressed Sensing}
\acrodef{cdf}[CDF]{Cumulative Distribution Function}
\acrodef{cef}[CEF]{Channel Estimation Field}
\acrodef{cacfar}[CA-CFAR]{Cell-Averaging Constant False-Alarm Rate}
\acrodef{cacc}[CACC]{Cross-Antenna Cross-Correlation}
\acrodef{casr}[CASR]{Cross-Antenna Signal Ratio}
\acrodef{cfo}[CFO]{Carrier Frequency Offset}
\acrodef{cir}[CIR]{Channel Impulse Response}
\acrodef{cfr}[CFR]{Channel Frequency Response}
\acrodef{csi}[CSI]{Channel State Information}
\acrodef{csirs}[CSI-RS]{CSI-Reference Signal}
\acrodef{cv}[CV]{Constant Velocity}
\acrodef{cnn}[CNN]{Convolutional Neural Network}
\acrodef{cots}[COTS]{Commercial-Off-The-Shelf}
\acrodef{dl}[DL]{Deep Learning}
\acrodef{dma}[DMA]{Direct Memory Access}
\acrodef{dmg}[DMG]{Directional Multi Gigabit}
\acrodef{dti}[DTI]{Data Transfer Interval}
\acrodef{dft}[DFT]{Discrete Fourier Transform}
\acrodef{dtft}[DTFT]{Discrete-Time Fourier Transform}
\acrodef{edmg}[EDMG]{Enhanced Directional Multi Gigabit}
\acrodef{ekf}[EKF]{Extended Kalman Filter}
\acrodef{elu}[ELU]{Exponential-Linear Unit}
\acrodef{fmcw}[FMCW]{Frequency-Modulated Continuous-Wave}
\acrodef{fov}[FOV]{Field-of-View}
\acrodef{ft}[FT]{Fourier Transform}
\acrodef{fo}[FO]{Frequency Offset}
\acrodef{fpga}[FPGA]{Field Programmable Gate Array}
\acrodef{gpio}[GPIO]{General Purpose Input/Output}
\acrodef{gsps}[GSPS]{Giga-Samples per Second}
\acrodef{gps}[GPS]{Global Positioning Systems}
\acrodef{har}[HAR]{Human Activity Recognition}
\acrodef{ht}[HT]{High Throughput}
\acrodef{ica}[ICA]{Independent Component Analysis}
\acrodef{if}[IF]{Intermediate Frequency}
\acrodef{ifs}[IFS]{Inter-Frame Spacing}
\acrodef{iht}[IHT]{Iterative Hard Thresholding}
\acrodef{imu}[IMU]{Inertial Measurement Unit}
\acrodef{isac}[ISAC]{Integrated Sensing And Communication}
\acrodef{isi}[ISI]{Inter-Symbol Interference}
\acrodef{idft}[IDFT]{Inverse Discrete Fourier Transform}
\acrodef{itu}[ITU]{International Telecommunication Union}
\acrodef{jcs}[JCS]{Joint Communication \& Sensing}
\acrodef{nnjpdaf}[NN-JPDAF]{Nearest-Neighbors Joint Probabilistic Data Association Filter}
\acrodef{los}[LoS]{Line-of-Sight}
\acrodef{lbm}[LBM]{Loop-Back Memory}
\acrodef{lte}[LTE]{Long-Term Evolution}
\acrodef{lo}[LO]{Local Oscillator}
\acrodef{mae}[MAE]{Mean Absolute Error}
\acrodef{mcs}[MCS]{Modulation and Coding Scheme}
\acrodef{md}[$\mu$D]{micro-Doppler}
\acrodef{mimo}[MIMO]{Multiple-Input and Multiple-Output}
\acrodef{mmwave}[mmWave]{Millimeter-Wave}
\acrodef{msps}[MSPS]{Mega-Samples per Second}
\acrodef{mu}[MU]{Multiple User}
\acrodef{music}[MUSIC]{MUltiple SIgnal Classification}
\acrodef{nac}[NAC]{Normalized Auto Correlation}
\acrodef{nco}[NCO]{Numerical Controlled Oscillator}
\acrodef{nlos}[NLoS]{Non-Line-of-Sight}
\acrodef{ofdm}[OFDM]{Orthogonal Frequency Division Multiplexing}
\acrodef{per}[PER]{Packet Error Rate}
\acrodef{phy}[PHY]{Physical Layer}
\acrodef{pl}[PL]{Programmable Logic}
\acrodef{pov}[POV]{Point-of-View}
\acrodef{pmn}[PMN]{Perceptive Mobile Network}
\acrodef{ps}[PS]{Processing System}
\acrodef{psdu}[PSDU]{Physical layer Service Data Unit}
\acrodef{rcs}[RCS]{Radar Cross Section}
\acrodef{rf}[RF]{Radio Frequency}
\acrodef{rfsoc}[RFSoC]{Radio Frequency System on a Chip}
\acrodef{rmse}[RMSE]{Root Mean Squared Error}
\acrodef{rss}[RSS]{Received Signal Strength}
\acrodef{rom}[ROM]{Read Only Memories}
\acrodef{sc}[SC]{Single Carrier}
\acrodef{sdr}[SDR]{Software Defined Radio}
\acrodef{siso}[SISO]{Single Input Single Output}
\acrodef{sls}[SLS]{Sector Level Sweep}
\acrodef{snr}[SNR]{Signal-to-Noise Ratio}
\acrodef{soc}[SoC]{System on a Chip}
\acrodef{spb}[SPB]{Signal Processing Blocks}
\acrodef{srrc}[SRRC]{Square-Root-Raised-Cosine}
\acrodef{ssb}[SSB]{Synchronization Signal Block}
\acrodef{ssr}[SSR]{Super Sample Rate}
\acrodef{sta}[STA]{Station}
\acrodef{stf}[STF]{Short Training Field}
\acrodef{stft}[STFT]{Short Time Fourier Transform}
\acrodef{su}[SU]{Single User}
\acrodef{tf}[TF]{Time-Frequency}
\acrodef{toa}[ToA]{Time of Arrival}
\acrodef{to}[TO]{Timing Offset}
\acrodef{usrp}[USRP]{Universal Software Radio Peripheral}
\acrodef{ue}[UE]{User Equipment}
\acrodef{vht}[VHT]{Very High Throughput}
\acrodef{wlan}[WLAN]{Wireless Local Area Network}
\acrodef{wvd}[WVD]{Wigner-Ville Distribution}